\documentclass[preprint,preprintnumbers,amsmath,amssymb,floatfix,nofootinbib]{article}
\pdfoutput=1
\usepackage{jheppub} 
\usepackage[utf8]{inputenc}
\usepackage{graphicx}
\usepackage{hyperref}
\usepackage{epsf, array, color}
\usepackage{graphicx}
\usepackage{bbold}
\usepackage{slashed}
\usepackage{amsthm}
\usepackage{amsmath}
\usepackage{multirow}
\usepackage{bm}
\usepackage{subcaption}

\DeclareMathAlphabet{\mathcalligra}{T1}{calligra}{m}{n}

\newcommand{\beq}{\begin{equation}}
\newcommand{\eeq}{\end{equation}}
\newcommand{\bal}{\begin{align}}
\newcommand{\eal}{\end{align}}
\newcommand{\bit}{\begin{itemize}}
\newcommand{\eit}{\end{itemize}}
\newcommand{\ben}{\begin{enumerate}}
\newcommand{\een}{\end{enumerate}}

\renewcommand{\eqref}[1]{Eq.~(\ref{eq:#1})}

\newcommand{\secref}[1]{Sec.~\ref{sec:#1}}

\newcommand{\figref}[1]{Fig.~\ref{fig:#1}}
\newcommand{\figsref}[2]{Figs.~\ref{fig:#1} and \ref{fig:#2}}

\newcommand{\tabref}[1]{Tab.~\ref{tab:#1}}

\newcommand{\f}{\frac}
\renewcommand{\d}{\text{d}}

\newcommand{\vev}[1]{ \left\langle {#1} \right\rangle }

\newcommand{\OO}{\mathcal{O}}

\newcommand{\cm}{\,\text{cm}}

\newcommand{\gev}{{\ \rm GeV}}
\newcommand{\tev}{{\ \rm TeV}}
\newcommand{\mev}{{\ \rm MeV}}

\newcommand{\be}{\begin{eqnarray}}
\newcommand{\ee}{\end{eqnarray}}
\newcommand{\bea}{\begin{eqnarray}}
\newcommand{\eea}{\end{eqnarray}}

\begin{document}  

\title{Exploring leptophilic dark matter with NA64-$\mu$}

\author[a,b]{Chien-Yi Chen,}
\affiliation[a]{Department of Physics and Astronomy, University of Victoria, Victoria, BC V8P 5C2, Canada}
\affiliation[b]{Perimeter Institute for Theoretical Physics, Waterloo, ON N2L 2Y5, Canada}
\emailAdd{cchen@perimeterinstitute.ca}

\author[c]{Jonathan Kozaczuk,}
\affiliation[c]{Amherst Center for Fundamental Interactions, Department of Physics,
 University of Massachusetts, Amherst, MA 01003, USA}
\emailAdd{kozaczuk@umass.edu}

\author[d]{Yi-Ming Zhong}
\affiliation[d]{Physics Department, Boston University, Boston, MA 02215, USA}
\emailAdd{ymzhong@bu.edu}

\abstract{We investigate the prospects for detecting light leptophilic dark sectors with a missing-momentum experiment at NA64 running in muon mode. In particular, we consider models in which dark matter connects to the visible sector through a lepton- or muon-specific scalar mediator. These scalars can also account for the $\sim 3.5\sigma$ discrepancy between the measured and predicted values of $(g-2)_{\mu}$. We emphasize the complementarity between NA64-$\mu$ and other terrestrial and astrophysical probes.}

\begin{flushright}
{\large 
ACFI-T18-15\\
}
\end{flushright}
\maketitle

\flushbottom

\setcounter{page}{2}

\section{Introduction}
\label{sec:intro}

The measured value of the anomalous magnetic moment of the muon, $a_\mu \equiv (g-2)_\mu/2$, differs from the Standard Model (SM) prediction~\cite{Olive:2016xmw} by
\be
\Delta a_\mu = a_\mu^{\rm EXP} -a_\mu^{\rm SM}=(268\pm 63 \pm 43) \times 10^{-11}.
\label{eq:g-2}
\ee
Here, the first and second error bars indicate the experimental and theoretical uncertainties, respectively. In terms of these uncertainties, the measured result represents a 3.5$\sigma$ upward deviation from the SM prediction. This discrepancy first surfaced about 15 years ago~\cite{Bennett:2004pv} and currently remains unexplained. On-going efforts to measure $(g-2)_\mu$ more precisely at Fermilab~\cite{Grange:2015fou,Anastasi:2015oea} and J-PARC~\cite{Mibe:2010zz}, along with improvements in the SM theoretical predictions from e.g.~advancements in lattice quantum chromodynamics (see~\cite{Nyffeler:2017ohp} for a recent review), may shed additional light on the discrepancy in the near future.  

At the same time, astrophysical and cosmological observations have provided strong evidence for the existence of dark matter (DM). To date, its identity remains undetermined and only its gravitational interactions have been observed. Null results at dark matter direct and indirect detection experiments and collider searches targeting $\sim \OO(10) \gev-\OO(10) \tev$ DM masses may be pointing towards scenarios beyond the typical weakly interacting massive particle (WIMP) paradigm. Light dark matter (LDM), with masses at the GeV scale or below, has been recognized as a particularly compelling and well-motivated alternative, receiving considerable attention in the literature and motivating several dedicated experimental efforts  (see~\cite{Alexander:2016aln,Battaglieri:2017aum} and references therein for an overview).  Typically, dark matter at or below the GeV scale requires an additional light particle to mediate interactions with the SM and avoid overproduction through thermal freeze-out in the early universe~\cite{Lee:1977ua,Boehm:2002yz,Boehm:2003hm,Pospelov:2007mp,ArkaniHamed:2008qn}. 

Light mediators, built into thermal LDM models, can also explain the $(g-2)_\mu$ anomaly. Several of the simplest scenarios furnishing new light degrees of freedom, such as dark photons and scalars mixing with the SM Higgs, are already disfavored as explanations of the $(g-2)_{\mu}$ anomaly by existing measurements (see e.g.~\cite{Krnjaic:2015mbs, Battaglieri:2017aum}). However, there remain several viable possibilities if the new mediator couples predominantly to leptons. These \emph{leptophilic} mediators will be our focus here.  

Arguably the most direct method for exploring new physics explanations of $(g-2)_{\mu}$ is through muon beams at accelerator facilities. Given the possible connection between dark matter and the $(g-2)_\mu$ anomaly, missing momentum searches at muon beam experiments are a particularly appealing possibility, since one might expect the mediator to decay invisibly to dark matter. It has been pointed out~\cite{Gninenko:2014pea, Krasnikov:2017sma, Gninenko:2018tlp} that the NA64 experiment can run in muon mode (dubbed NA64-$\mu$) and perform a muon missing momentum search with a $\mu^+$ beam\footnote{In what follows, we refer to a $\mu^+$ beam as a ``muon beam" or ``$\mu$-beam" for simplicity unless a distinction between $\mu^+$ and $\mu^-$ is necessary.} supplied by the Super Proton Synchrotron (SPS) at CERN~\cite{adeva1994measurement,spsweb}.  If approved, NA64-$\mu$ is expected to run after the CERN long shutdown (2021). More recently, M$^3$, a compact muon missing momentum search experiment at Fermilab, has been proposed~\cite{Kahn:2018cqs}. It aims at the same measurement with potentially more muons on target (MOT). 

In this study, we consider leptophilic dark matter models with light scalar mediators as physics targets for these experiments, focusing in particular on the prospects for NA64-$\mu$. We will be interested in models in which the mediator decays primarily to dark matter, and hence evades most searches for visibly-decaying light particles. In contrast to the gauged $L_\mu-L_\tau$ scenarios often discussed in the literature, light scalar mediators need not couple to neutrinos and can feature substantial freedom in the couplings to different lepton flavors. In fact, the couplings to all but one flavor could be strongly suppressed~\cite{Batell:2017kty}. These models thus represent a distinct class of viable but experimentally challenging explanations of both DM and the $(g-2)_\mu$ anomaly. We will show that the parameter space of these models resolving the $(g - 2)_\mu$ discrepancy can be conclusively probed with NA64-$\mu$ for mediator masses below $\sim 10$ GeV. We will also emphasize the complementarity between NA64-$\mu$ and other existing and proposed probes of light leptophilic dark sectors with invisibly-decaying mediators. We believe our results strengthen the scientific case for experiments like NA64-$\mu$ and provide additional well-motivated physics targets that would be difficult to experimentally access otherwise.

Before proceeding, let us briefly comment on the relation of the present study to previous work appearing in the literature. The prospects for NA64-$\mu$ in exploring leptophilic vector bosons were studied in Refs.~\cite{Gninenko:2014pea, Gninenko:2018tlp} in the context of gauged $L_\mu - L_\tau$ models.  We utilize and extend their NA64 sensitivity estimates to the models introduced below. Ref.~\cite{Kahn:2018cqs} discussed M$^3$ projections and constraints primarily in terms of a gauged $L_\mu - L_\tau$ dark matter model. The authors also consider detection prospects for invisibly-decaying scalars coupled to muons at M$^3$, and we utilize their results when comparing NA64-$\mu$ against M$^3$. Refs.~\cite{Batell:2016ove, Chen:2015vqy} explored explanations of the $(g-2)_\mu$ discrepancy with light leptophilic scalars coupling to all lepton flavors and decaying visibly. The leptophilic model we consider resembles those considered in these studies, extended to include couplings to dark matter.  Ref.~\cite{Knapen:2017xzo} studied light dark matter models with scalar mediators coupled to electrons, and we incorporate some of their results in discussing the cosmological and astrophysical signatures of our models. Ref.~\cite{Batell:2017kty} introduced muon-specific scalar models and discussed ultraviolet (UV) completions of these scenarios, again focusing on scalars that decay visibly. We make use of their results in motivating models with muon-specific mediators. Finally, Ref.~\cite{Chen:2017awl} considered light leptophilic and muon-specific mediators and their detection at NA64-$\mu$, focusing on the case with visible (but displaced) $S$ decays. Our study can be viewed as an extension of this work to the case where the scalar mediates interactions of a dark sector with the SM and decays invisibly.

The remainder of this study is structured as follows. In~\secref{models}, we introduce light dark matter models with leptophilic scalar mediators and discuss the associated cosmological and astrophysical consequences.~\secref{NA64} discusses the muon missing momentum search at NA64-$\mu$ and its prospects for exploring these models. In~\secref{complementarity} we survey several other probes of these models and compare them to the sensitivities afforded by NA64-$\mu$. We conclude in~\secref{Conclusion}. 

\section{Leptophilic dark matter with scalar mediators}
\label{sec:models}

\subsection{Model setup and general considerations}
Motivated by the long standing $(g-2)_\mu$ anomaly, we consider dark sectors with leptophilic interactions with the Standard Model. For concreteness, we will take the dark matter candidate to be a Dirac fermion $\chi$, and the scalar mediator, $S$, to only couple to SM leptons. In UV-complete models, one might also expect $S$ couplings to quarks. Our setup should be understood as corresponding to models in which the scalar's couplings to quarks are small relative to its couplings to leptons, so that the latter dominate the phenomenology. Given these assumptions, the effective Lagrangian governing the interactions of the scalar with the SM leptons and a Dirac fermion dark matter candidate is taken to be
\begin{align}\label{eq:L}
\mathcal L \supset{}& - g_\chi S \bar{\chi} \chi - \sum_{\ell = e,\mu,\tau} g_\ell S \bar{\ell} \ell.
\end{align}
We will assume that $S$ and $\chi$ are both light, with masses below $\OO(10)$ GeV, and take $S$ to be \emph{real}. 

The couplings of $S$ to leptons in~\eqref{L} violates the $SU(2)_L\times U(1)_Y$ gauge invariance of the Standard Model, but can be understood as originating from the effective gauge-invariant dimension-5 operators
\begin{equation} \label{eq:EFT}
\frac{c_i}{\Lambda} S\overline{L}_iH E_i.
\end{equation}
Here $\Lambda$ is the associated scale of new physics and $c_i$ is a Wilson coefficient for the flavor $i$. We will assume that the couplings are diagonal in the mass basis. While the relative sizes of the Wilson coefficients $c_i$ (and hence the effective couplings $g_i$) are undetermined \textit{a priori}, a natural expectation might be that they are proportional to the corresponding Yukawa coupling $y_i$, so that the effective $g_i$ are proportional to the corresponding lepton masses after electroweak symmetry breaking. This is the case in the framework of Minimal Flavor Violation (MFV), for example. It is also possible, however, to have new physics at the scale $\Lambda$ with non-minimal flavor structure. As emphasized in~\cite{Batell:2017kty}, this could give rise to couplings of $S$ predominantly to one flavor in a technically natural way that avoids dangerous flavor changing neutral currents. In our analysis below, we will consider both the MFV-motivated lepton-specific case, with mass-proportional couplings, and the muon-specific case, in which the couplings to electrons and taus are negligible, i.e.,
\begin{equation}
\label{med}
\text{Scalar mediator ($S$) =}
  \left \{
  \begin{tabular}{l}
     Lepton-specific scalar: \quad $g_e:g_\mu:g_\tau = m_e: m_\mu : m_\tau$\\
      Muon-specific scalar: \quad $g_\mu \neq 0,\, g_e = g_\tau = 0.$
  \end{tabular}
\right.
\end{equation}

Throughout our study we will remain agnostic about the particular UV completion of the effective operators in Eq.~(\ref{eq:L}), focusing on the model-independent constraints and prospects for observation. Possible UV completions involving lepton- or muon-specific scalars have been proposed in the literature and include scenarios with new vector-like leptons and lepton-specific two-Higgs-doublet plus singlet models (see e.g.~\cite{Batell:2016ove, Chen:2015vqy, Batell:2017kty}). Adding a coupling of the scalar to dark matter in most of these models is trivial. From the EFT perspective, assuming  Wilson coefficients proportional to the corresponding Yukawa couplings, $c_{\ell} \sim \mathcal{O}(1)\times y_{\ell}$ in Eq.~(\ref{eq:EFT}), new physics scales $\Lambda \gtrsim 1$ TeV correspond to muon couplings $g_{\mu} \lesssim \mathcal{O}(10^{-4}-10^{-3})$. For low values of $m_S$, we will see that the $a_\mu$-favored region falls in this regime, as can the thermal relic target for leptophilic dark matter. If the new physics responsible for generating the operators in Eq.~(\ref{eq:EFT}) does not involve new colored states, the LHC is unlikely to constrain the corresponding UV completions for $\Lambda$ near the TeV scale. Couplings $g_\mu \gtrsim \mathcal{O}(10^{-3})$ correspond to $\mathcal{O}(100$ GeV$)$ scales of new physics (assuming $c_{\ell} \sim \mathcal{O}(1)\times y_{\ell}$ ), and so the EFT can break down at LHC energies and a UV completion should be specified when considering constraints at high-energy experiments. Nevertheless, since several UV completions have been shown to remain viable while generating couplings in this range~\cite{Chen:2015vqy, Batell:2016ove, Batell:2017kty}, and the Wilson coefficients need not be proportional to the corresponding Yukawa coupling, we content ourselves with the UV-agnostic treatment in what follows. Note also that, as discussed in Ref.~\cite{Batell:2017kty}, the scenarios we consider can be consistent with technical naturalness provided new physics enters to cut off the quadratically-divergent contributions to $m_S^2$ around the TeV scale.

The $S$-lepton couplings in~\eqref{L} also introduce effective scalar couplings to pairs of vector bosons ($SVV$) at one-loop. For a light scalar with mass below the electroweak (EW) scale, the most relevant $SVV$ coupling is the scalar-diphoton coupling ($S\gamma\gamma$).\footnote{Couplings like $g_{Z\gamma}$ can be also relevant. We discuss the corresponding signatures in Sec.~\ref{sec:complementarity}.} This interaction is important when the decays of $S\to \ell^+ \ell^-$ are kinematically forbidden and is accounted in our thermal relic density calculations below.  Additionally, the $S\gamma \gamma$ coupling can give rise to experimental signatures involving photons, also discussed below. We can parametrize the corresponding interactions with an effective Lagrangian,
\begin{align}\label{eq:Lphoton}
\mathcal L \supset{}& - \frac{1}{4} g_{\gamma\gamma} S F_{\mu \nu} F^{\mu \nu}, \quad 
g_{\gamma\gamma} = \frac{\alpha}{2 \pi} \left|\sum_{\ell = e, \mu, \tau} \f{g_\ell}{m_\ell}F_{1/2} \left(\f{4 m_\ell^2}{p_S^2},\f{q^2}{4m_\ell^2}\right) \right|,
\end{align}
where $\alpha$ is the electromagnetic fine structure constant and $F_{1/2}$ is a form factor that depends on the four-momentum-square of one of the photons ($q^2$) and the scalar $S$ ($p_S^2$). In computing $F_{1/2}$ we will consider one of the photons to be on-shell but allow $S$ and the other photon (with four-momentum $q$) to be off-shell. Expressions for $F_{1/2}$ are given in Appendix~\ref{sec:app}. Combining~\eqref{L} and~\eqref{Lphoton}, the total decay width for $S$ is given by
\beq
\Gamma_S =\sum_{i = e, \mu, \tau, \chi} \f{g_i^2 m_S}{8\pi} \left(1-\f{4m_i^2}{m_S^2}\right)^{3/2}+\f{\alpha^2 m_S^3}{256\pi^3} \left|\sum_{\ell = e, \mu, \tau}\f{g_\ell}{m_\ell} F_{1/2}\left(\frac{4 m_\ell^2}{m_S^2},0\right)\right|^2.
\label{eq:width}
\eeq
We will primarily focus on cases where $g_\chi \gg g_\ell$, so that $S$ decays primarily to $\chi \overline \chi$. For the decay of $S$ to two photons, all three particles are on-shell, corresponding to $q^2=0$ and $p_S^2=m_S^2$ in $F_{1/2}$.

As emphasized in~\secref{intro}, besides mediating interactions between the visible and hidden sectors, the scalar $S$ can contribute to $(g-2)_\mu$ through its couplings to muons ~\cite{Leveille:1977rc, Lindner:2016bgg}:
\beq
\Delta a_{\mu} = \frac{g_\mu^2}{8\pi^2}\int_0^1 dz \frac{(1-z)^2(1+z)}{(1-z)^2+z(m_S/m_{\mu})^2}.
\eeq
This contribution is positive, and can raise the predicted value of $a_\mu$ so that it agrees with experiment, cf.~\eqref{g-2}. In our discussion of the model parameter space below, we will indicate the regions consistent with the central $a_\mu$ value within 2$\sigma$, as well as regions expected to be favored by future $a_\mu$ measurements, assuming that the central value of $\Delta a_\mu$ will remain unchanged while the experimental and theoretical uncertainties will be improved by a factor of 4 and 2, respectively \cite{Olive:2016xmw,Anastasi:2015oea,Grange:2015fou, Mibe:2010zz, Jegerlehner:2018zrj}.

Turning our attention to the DM, there are three distinct possibilities for the relative sizes of $m_\chi$ and $m_S$ that carry different phenomenological consequences:
\begin{itemize}
\item  $m_\chi < m_S/2$: in this case, for $g_{\chi} \gtrsim g_{e, \mu, \tau}$, the mediator $S$ will primarily decay invisibly to $\chi \overline{\chi}$.  The thermal freeze-out relic abundance of DM is driven by $s$-channel annihilation into leptons, $\overline\chi \chi \to \overline\ell \ell$ (or $\gamma\gamma$) in the Early Universe. The annihilation rate is roughly given by  
\beq 
\vev{\sigma v} =\f{1}{8\pi} {g_D^2 g_{\ell}^2 \over m_\chi} {(m_\chi^2-m_{\ell}^2)^{3/2} \over (m_S^2-4 m_\chi^2)^2}\langle v_{\rm rel}^2\rangle, \label{eq:relic}
 \eeq 
 and depends on both the dark sector coupling $g_{\chi}$ and the visible sector coupling(s) $g_{\ell}$ (a sum over lepton flavors is implicit above). $\langle v_{\rm rel}^2\rangle$ is the thermal average of the relative DM velocity squared, and its presence above reflects the fact that annihilation is a $p$-wave process in this scenario. Since the annihilation rate depends on $g_\ell$, this scenario provides a well-defined thermal dark matter target that can be searched for in terrestrial experiments. 
      
\item $m_S/2 \lesssim m_\chi \lesssim m_S$: here again $s$-channel annihilation into leptons sets the relic abundance of $\chi$, providing a thermal relic target. However, $S\to  \chi \overline \chi$ decays are kinematically forbidden, and so $S$ will decay visibly. This dramatically changes the constraints and prospects for detection at accelerator experiments, and has been discussed in detail elsewhere in the literature (see e.g.~\cite{Chen:2015vqy,Batell:2016ove}). In particular, the NA64-$\mu$ projections of interest below can be taken from Ref.~\cite{Chen:2017awl}.

\item $m_S < m_\chi$: in this case, again $S$ decays visibly. However, annihilation in the early Universe will primarily proceed through secluded annihilation, $\overline \chi \chi \to S S$. The cross-section for this process only depends on the dark sector coupling $g_{\chi}$, and so there is no well-defined thermal relic target for terrestrial experiments. Nevertheless, this is a viable possibility, and again the prospects for discovery can be inferred from e.g.~\cite{Batell:2016ove, Chen:2017awl}.
\end{itemize}
 In what follows, we will focus on the first of these scenarios, with $m_\chi < m_S/2$, since it provides concrete thermal targets and is generally the most difficult to test, given the invisible decays of $S$. 

\subsection{Relic abundance via thermal freeze-out}

To begin our investigation of the parameter space of these models, we first determine the regions consistent with the observed relic abundance of dark matter. We assume the dark sector is in thermal equilibrium with the SM plasma in the Early Universe~\cite{Knapen:2017xzo}. At later times, the annihilation rate for $\overline\chi \chi \to \overline\ell \ell$ (or $\gamma\gamma$) drops below the expansion rate of the Universe and the dark matter abundance freezes out. To accurately compute the resulting relic abundance, $\Omega_{\chi}h^2$, we use the \texttt{MadDM 2.1}~\cite{Ambrogi:2018jqj} incorporating a UFO-format model built in \texttt{FeynRules 2.3.27}~\cite{Alloul:2013bka}. For concreteness, we compute $\Omega_{\chi}h^2$ for two specific DM--mediator mass ratios, 
\begin{align}
m_\chi ={}& \frac{m_S}{3},
\label{eq:kin1}\\
m_\chi ={}& \f{m_S}{2 \sqrt{1+\epsilon_R}}\approx \frac{m_S}{2}\left(1-\frac{\epsilon_R}{2}\right),
\label{eq:kin2}
\end{align}
where we assume $\epsilon_R \ll 1$, adopting the notation from~\cite{Feng:2017drg}. The second mass ratio is chosen to illustrate the effects of the resonant enhancement present for the $s$-channel annihilation process when $m_{\chi}\sim m_S/2$. As emphasized in Ref.~\cite{Feng:2017drg}, this enhancement can dramatically affect the thermal target parameter space, allowing for smaller couplings consistent with the observed dark matter density, $\Omega_{\rm DM}h^2 \approx 0.12$~\cite{Ade:2015xua}. Results for the mass relations in~\eqref{kin1} (\eqref{kin2}) in the lepton-specific and muon-specific models are shown in the left and right panels of Fig.~\ref{fig:money_plot} (\figref{money_plot_2}), respectively. Regions below the red contours are excluded due to an over-abundance of DM for the indicated mass ratio. ``Kinks" around $m_S = 2m_\mu$ (both left and right panels) and $m_S = 2 m_\tau$ (right panel) occur when a new annihilation channel becomes kinematically accessible as indicated by~\eqref{relic}. The contrast between the relic density-compatible regions in~\figref{money_plot} and~\ref{fig:money_plot_2} illustrates the model-dependence of the thermal freeze-out constraint. Figs.~\ref{fig:money_plot} and \ref{fig:money_plot_2} also show the corresponding $a_\mu$-favored regions in the $m_S-g_\mu$ parameter space. Comparing the relic density curves with the green bands suggests that, given a suitable choice for $m_S/m_\chi$, a large portion of the parameter space favored by the $(g-2)_\mu$ measurement can also furnish a potentially viable dark matter candidate (which may or may not saturate the entire observed abundance).

\subsection{Constraints on $g_\chi$ from dark matter self-interactions and perturbativity}

\begin{figure}[t]
   \centering
   \includegraphics[width=0.6\textwidth]{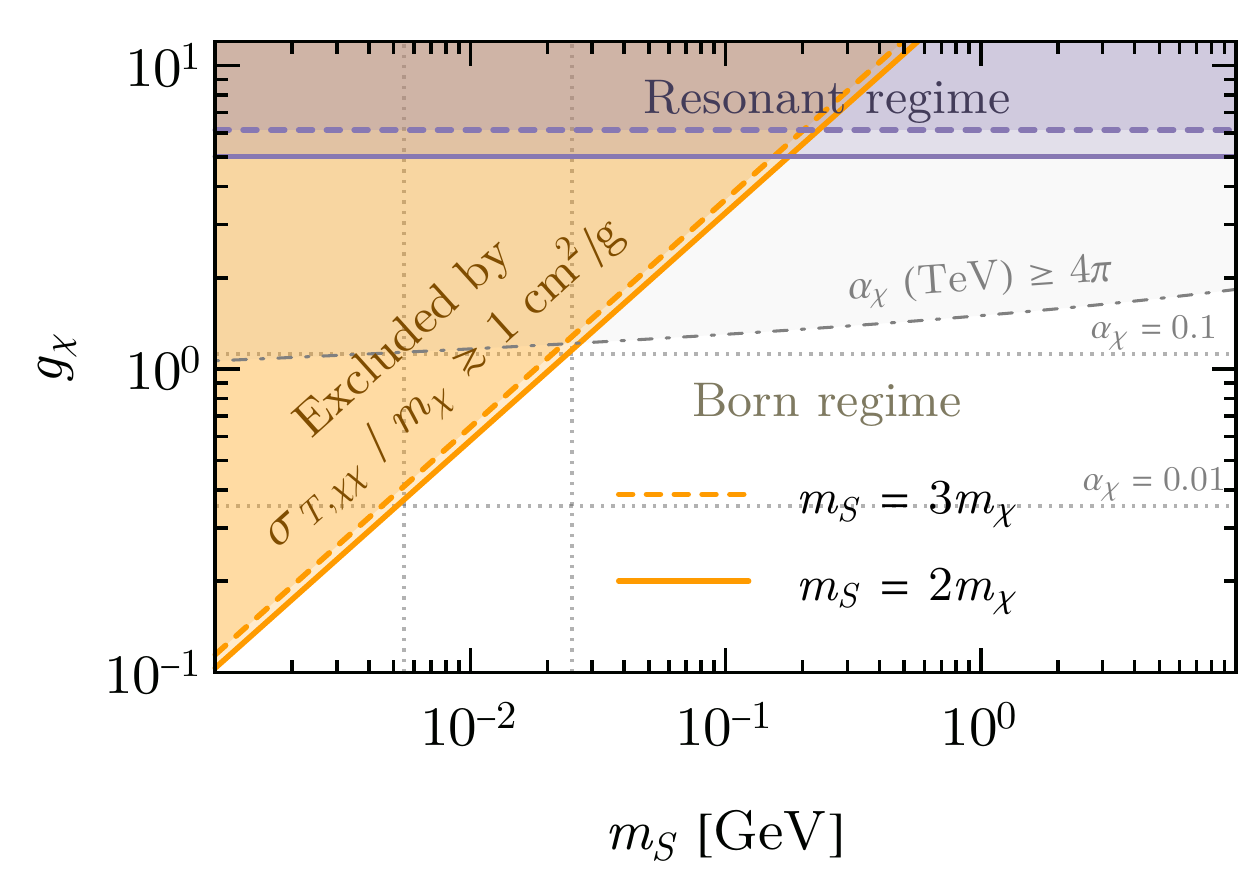}
   \caption{Constraints on $g_\chi$ for $m_\chi = m_S/3$ (dashed) and $m_\chi = m_S/(2\sqrt{1+\epsilon_R}) \approx m_S/2$ (solid) from DM self-interactions and perturbativity requirements. The orange region is disfavored by astrophysical constraints such as halo morphism (see discussions in the text). The purple region indicates the resonant regime ($\alpha_\chi\equiv g_\chi^2/4\pi \gtrsim m_S /m_\chi$ and $m_\chi v_\text{rel} /m_S \lesssim 1$) where~\eqref{selfinteraction} is not applicable. Above the gray dash-dotted line, $\alpha_\chi$ runs to a value $ \geq 4\pi$ at the TeV scale or below.}
   \label{fig:selfint}
\end{figure}

The presence of the scalar mediator also introduces a Yukawa-like attractive self-interaction between DM particles~\cite{Tulin:2013teo}. Such interactions can be constrained by astrophysical observations. Taking $m_\chi < m_S/2$ and given that $v_\text{rel} \lesssim \OO(10^{-3})$ in most astrophysical systems, the resulting momentum-transfer self-interaction cross-section strength\footnote{Here we use the momentum-transfer cross-section (a.k.a.~the diffusion cross-section) rather than the viscosity  cross-section in order to compare results from N-body simulations such as~\cite{Peter:2012vi}. } 
is given by
\beq
\f{\sigma_{T, \chi\chi}}{m_\chi} =\f{g_\chi^4 m_\chi}{4\pi m_S^4} = 1.7\times 10^{-5}\,\text{cm}^2/\text{g}  \left(\f{g_\chi}{1}\right)^4 \left(\f{m_\chi}{1 \gev}\right) \left(\f{1 \gev}{m_S}\right)^4,
\label{eq:selfinteraction}
\eeq
if the interaction falls in the perturbative Born regime, $\alpha_\chi \equiv g_\chi^2/4\pi \ll m_S /m_\chi$ (as well as $m_\chi v_\text{rel} /m_S \lesssim 1$). 

Given that~\eqref{selfinteraction} is velocity-independent, self-interaction constraints from halo systems at different scales should be taken into account simultaneously. On the dwarf galaxy scale,  $\sigma/m \lesssim \OO(10\,\text{cm}^2/\text{g})$ is allowed since gravothermal collapse of the halo is avoided~\cite{Balberg:2002ue, Koda:2011yb}. For Milky Way--sized galaxies, $\sigma/m \lesssim 1\,\text{cm}^2/\text{g}$ is allowed by halo morphism~\cite{Peter:2012vi}. At the galaxy cluster scale, $\sigma/m  \lesssim 0.7-7\,\text{cm}^2/\text{g}$ is allowed by halo mergers (see e.g.~\cite{Tulin:2017ara}). Given these considerations, we choose $1\,\text{cm}^2/\text{g}$ as a robust upper limit on $\sigma_{T,\chi\chi}/m_\chi$. The resulting constraint is shown in~\figref{selfint}. As expected from~\eqref{selfinteraction}, the constraint on $g_\chi$ weakly depends on $m_\chi$ (or the ratio of  $m_\chi/m_S$). We see that $g_\chi =1\,(0.1)$ is disfavored for $m_S < 20\mev\, (1\mev)$ and not constrained for larger scalar masses. 

Large values of $g_\chi$ can also lead to a breakdown of perturbation theory at low scales, rendering perturbative results invalid. In \figref{selfint}, we also indicate regions for which $\alpha_\chi \geq 4\pi$ at the TeV scale or below. These results make use of the one-loop beta function for a real scalar coupled to a Dirac fermion~\cite{Cheng_1974, cline:1998} \beq \beta_{g_\chi} \equiv \f{\d}{\d \ln \mu} g_\chi = \f{5 }{16\pi^2} g_\chi^3,\eeq where $\mu$ is the renormalization scale. Motivated by the self-interaction and perturbativity constraints in ~\figref{selfint}, we take $g_\chi =1$ and restrict the scalar mass to be above 20 MeV in what follows.

\subsection{Other cosmological and astrophysical constraints}

There are other astrophysical and cosmological constraints on light leptophilic dark sectors. Observations of the cosmic microwave background (CMB) constrain the amount of energy injected into the Intergalactic Medium (IGM) through dark matter annihilation at late times. This energy injection can distort the CMB. However, this is only an issue if the DM annihilation cross-section is $s$-wave~\cite{Ade:2015xua,deNiverville:2012ij,Slatyer:2015jla}. Scenarios with a light scalar mediator and Dirac fermion dark matter feature a $p$-wave annihilation cross-section and therefore the CMB constraint is not relevant for the scenarios we consider.
Light dark matter can also affect the successful predictions of big bang nucleosynthesis (BBN). If dark matter is sufficiently light, it is expected to have been in equilibrium with the SM thermal bath, and hence relativistic, until after the onset of BBN. This affects the Hubble rate, and thus the abundances predicted by BBN, which are tightly constrained. For dark matter masses above about 1 MeV, this is not an issue, as freeze-out occurs sufficiently early, and thus this constraint does not significantly impact the scenarios we consider.

Dark scalars with mass below $ \OO (100\mev) $ can also be abundantly produced in a core-collapsed supernova (SN) through resonant or continuum production~\cite{Hardy:2016kme,Batell:2017kty, Knapen:2017xzo}. Once produced, $S$ can provide a new cooling channel for the supernova as it streams out from the core, and could conflict with observations of SN 1987A. We will discuss the corresponding SN 1987A constraint in~\secref{supernova}.

\section{Missing momentum measurements at NA64-$\mu$ } \label{sec:NA64}

\renewcommand{\d}{\text{d}}

We have argued that light leptophilic dark sectors are compelling from the standpoint of explaining both the observed DM abundance and the $(g-2)_\mu$ discrepancy. How might one explore these scenarios experimentally? The NA64-$\mu$ experiment was proposed in~\cite{Gninenko:2001hx, Gninenko:2014pea} and is currently planned to run after 2021. It was originally envisioned to search for new vector gauge bosons solving the $(g-2)_\mu$ puzzle. In this section, we study the possibility of using this experiment to search for light dark matter produced through the decay of a scalar mediator in the class of models motivated and described above.  We show that NA64-$\mu$ will provide impressive experimental coverage of these scenarios that are otherwise difficult to probe. 

\subsection{Production of light dark matter}
\label{sprod}

\begin{figure}[!t]
    \centering
    \begin{subfigure}[b]{0.33\textwidth}
        \includegraphics[width=\textwidth]{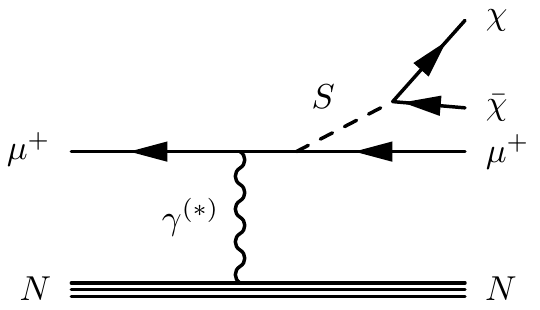}
        \caption{}
        \label{fig:schichi}
    \end{subfigure}
    \quad\quad 
    \begin{subfigure}[b]{0.33\textwidth}
        \includegraphics[width=\textwidth]{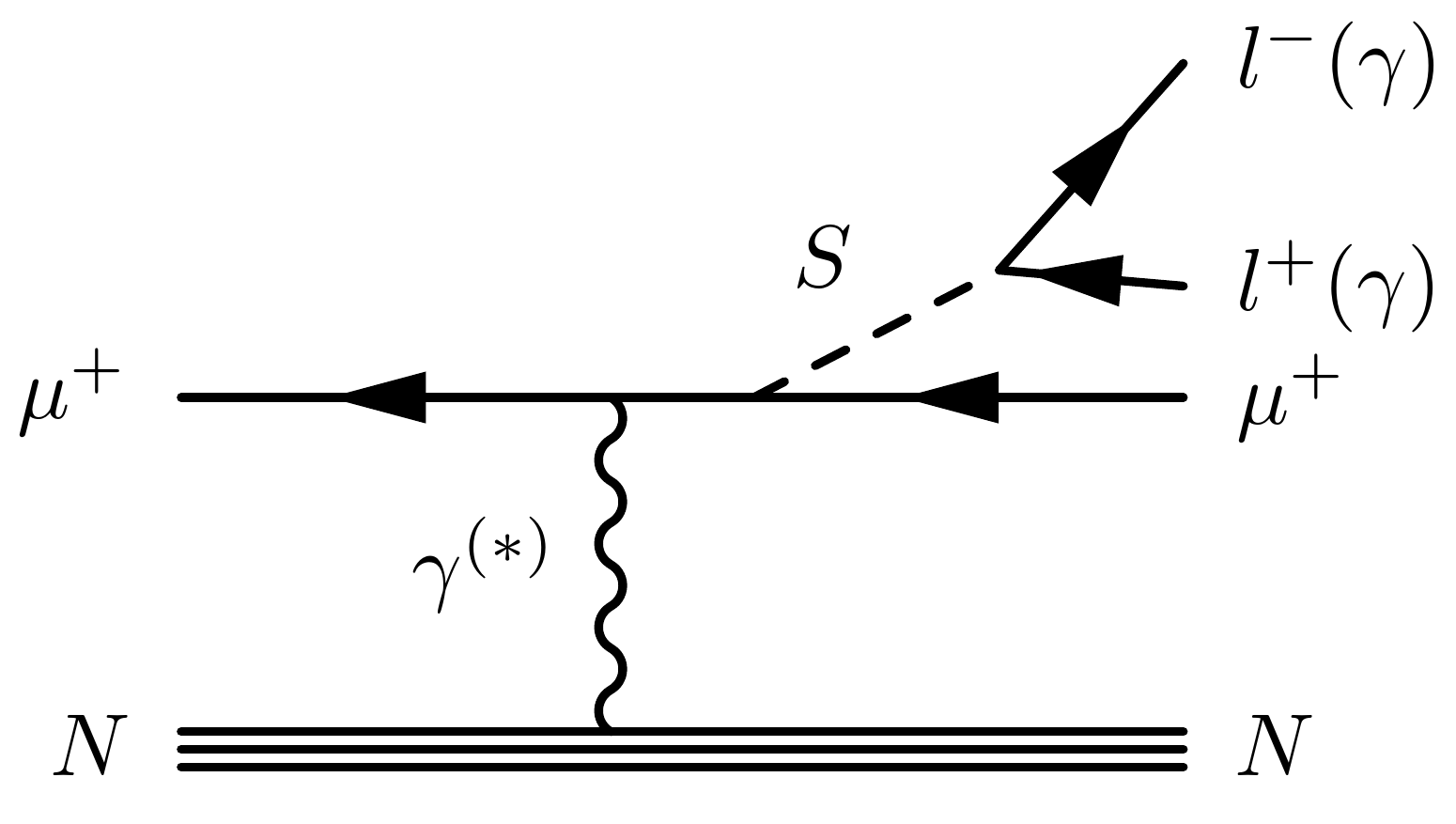}
        \caption{}
        \label{fig:sll}
    \end{subfigure}
    \caption{Diagrams contributing to the signal in a muon missing momentum search for dark scalar bremsstrahlung (a) with the scalar decaying invisibly and (b) the scalar decaying visibly but outside the detector. Given the NA64-$\mu$ setup and the parameter space we interested in, the missing momentum signals in our models are dominated by (a).}\label{fig:signaldiagram}
\end{figure}

At muon-beam experiments, dark scalars, $S$, can be produced as initial and finial state radiation off of the incident muons when they scatter with the target nuclei ($\mu^+ N \to \mu^+ N S$). Through this mechanism, and depending on the mass of $S$ and its couplings to dark matter and the SM leptons, a missing momentum measurement at NA64-$\mu$ would be capable of probing both (a) invisible decays of $S$ and (b) visible decays of $S$ outside of the detector, illustrated in~\figref{signaldiagram}. In both cases, dark scalar bremsstrahlung can induce a significant amount of missing energy/missing momentum\footnote{In our scenarios, DM interactions with the calorimeter material are expected to be negligible.} and leave a feebly scattered muon as a distinct signal.\footnote{Since the momentum (energy) of the outgoing muon can be measured (inferred) from the recoil trackers, respectively, there is no material distinction between a missing-momentum and a missing-energy measurement in this context, unlike at electron-beam experiments.} In the scenarios of interest,  
the expected number of events corresponding to case (a) is much greater than that from case (b) for the following reasons: (1) large values of $g_\chi$ are favored from the standpoint of obtaining the correct relic density without fine-tuning, while large values of $g_\ell$ are significantly constrained for the parameter space we are interested in. Thus, we expect BR$(S\to \chi\bar{\chi}) >$ BR$(S\to \ell\bar{\ell})$. (2) Case (b) surfers a exponential decay volume suppression in order for $S$ to decay outside the detector. There is no such suppression for case (a).  

A detailed calculation of the dark scalar bremsstrahlung ($\mu^+  N \to \mu^+ N  S$) production rate is discussed in \cite{Chen:2017awl}. We will briefly review it here and adopt it to estimate the number of missing momentum signal events expected in our models. The calculation is based on the improved Weizsacker-Williams (IWW) approximation~\cite{Kim:1973he}, which treats the exchanged virtual photon between the muon and nucleon as a real photon. It is a good approximation when the beam energy is much larger than the momentum transfer~\cite{Liu:2016mqv}.\footnote{Note that for lower dark scalar masses, non-negligible differences between the cross sections resulting from the IWW approximation and those from exact tree-level calculations might be expected~\cite{Gninenko:2017yus}.}  The total number of missing momentum events, $N_\chi$, can be approximated as
\begin{align}
N_\chi \simeq  \frac{ N_\mu n_{\rm atom}}{\langle \d E_\mu/\d y \rangle} \int_{E_{\mu, \min}}^{E_{\mu, \text{beam}}} \d E_\mu  \int_{x_{\min}}^1 \d x  {\d\sigma_{\mu N \to \mu N S} \over \d x}  \text{BR}({S\to \chi \bar{\chi}}),
\label{eq:Ns}
\end{align}
assuming that ${\d E_\mu/\d y}$, the change of the muon energy $E_\mu$ with respect to the penetration length $y$, is approximately constant. For the lead (Pb) target used in NA64-$\mu$, this is a good approximation for the relevant energy range ($\sim 100 \gev$), with $\langle \d E_\mu/\d y \rangle\approx 12.7\times 10^{-3}$ GeV/cm~\cite{pdgweb}.  In~\eqref{Ns},  $N_\mu$ is the total number of incident muons and $n_\text{atom}$ is the atomic number density of the target ($n_{\rm atom}=3.3\times 10^{22}/\cm^3$ for lead). The integration is over the penetrating muon energy $E_\mu$ and the bremsstrahlung scalar energy $E_S$ with respect to $E_\mu$ ($x\equiv E_S/E_\mu$). The lower integration limit for $E_\mu$, $E_{\mu,\min}  = E_{\mu, \text{beam}} - L_\text{tg} \langle \d E_\mu/\d y \rangle $, is set by the energy loss of a positive muon after passing through the entire target of length $L_\text{tg}$ (in the projection below, we use a thin target with $L_\text{tg}=20$ cm).
$E_{\mu, \text{beam}}$, the initial muon beam energy, is 150 GeV.
The lower integration limit on $x$, $x_{\min}$, is set by requirements for background rejection. Here, we choose $x_{\min}= 1/3$ as suggested by~\cite{Gninenko:2014pea}. This amounts to requiring signal events to have missing energy larger than $E_\mu/3\sim 50 \gev$. 

In most of the parameter space under consideration,  $\text{BR}({S\to \chi \bar{\chi}})$ appearing in \eqref{Ns} is very close to one. Nevertheless, it can decrease substantially when visible decays are enhanced by either a large $g_\ell$ or large phase space factors, and we account for this in our sensitivity estimates. 

The differential signal production cross-section appearing in \eqref{Ns} is given by\footnote{There is a typo in Ref~\cite{Chen:2017awl}, where the pre-factor 1/(4$\pi$) of Eq. (12) should instead be 1/(12$\pi$).}:
\be
\f{\d \sigma_{\mu N\to \mu N S}}{\d x} \simeq \f{g_\mu^2 \alpha^2}{12\pi} \chi \beta_\mu \beta_S \f{x^3\left[m_\mu^2 (3x^2-4 x+4)+2 m_S^2 (1-x)\right]}{\left[m_S^2(1-x)+m_\mu^2 x^2\right]^2},
\ee
where the boost factors are $\beta_\mu = \sqrt{1-m_\mu^2/E_\mu^2}\approx 1$ and $\beta_S = \sqrt{1-m_S^2/(E_S)^2}$ for muons and $S$, respectively. The effective photon flux, $\chi$, is given by
\be
\chi = \int_{t_{\min}}^{t_{\max}} \d t  \f{t-t_{\min}}{t^2} G_2(t)\simeq \int_{m_S^4/\left(4 E^2_\mu\right)}^{m_S^2+m_\mu^2} \d t  \f{t-m_S^4/\left(4 E^2_\mu\right)}{t^2} G_2(t),
\label{chiex}
\ee
where the virtuality $t$ represents the momentum transfer squared and $G_2$ is the combined atomic and nuclear form factor. An explicit expression for $G_2$ can be found in  Appendix A of Ref.~\cite{Chen:2017awl}. Finally, \eqref{Ns} assumes $\sim$100\% trigger and reconstruction efficiencies. The reader should bear this is mind in what follows.

\subsection{Experimental setup}
\label{sec:setup}

\begin{figure}[!t]
   \centering
   \includegraphics[width=0.96\textwidth]{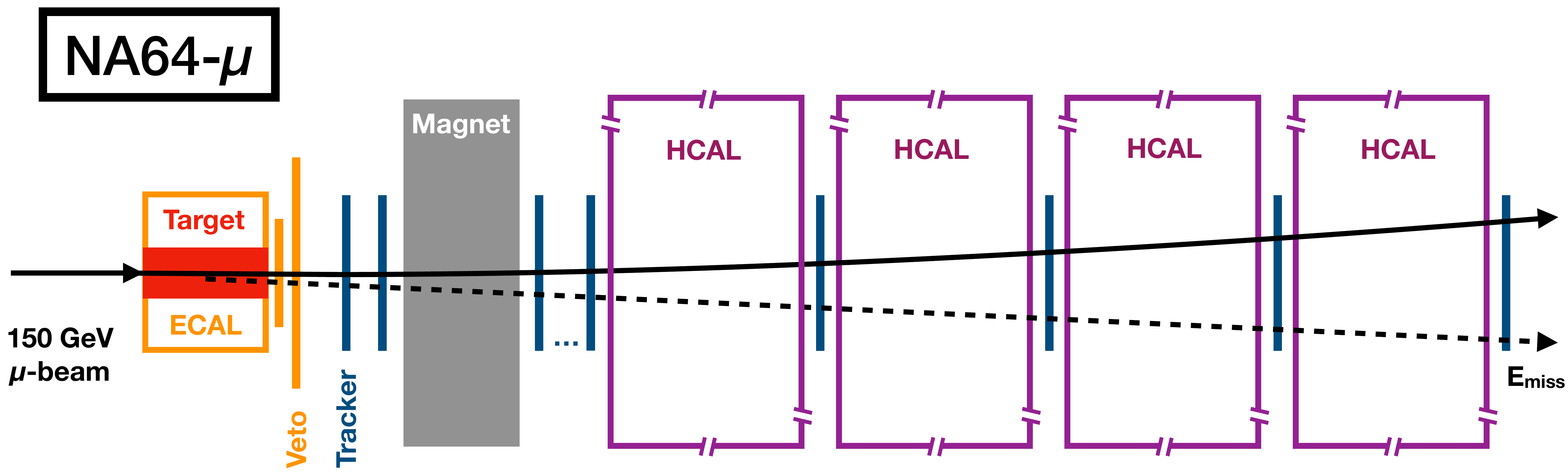} 
   \caption{The detector segment of the proposed NA64-$\mu$ experimental setup. See text for more details. The full configuration can be found in~\cite{Gninenko:2014pea, na64web}.}
   \label{fig:setupna64}
\end{figure}
With the signal rate computed, we can consider the sensitivity achievable by the NA64-$\mu$ experimental setup.  NA64-$\mu$ is equipped with a high energy and high intensity  muon-beam from the CERN SPS~\cite{adeva1994measurement,spsweb}. The muon beam has a maximum momentum  in the range of $100-225 \gev$. A typical intensity of $2\times 10^8$ $\mu^+$ per spill can be achieved for beam energies between 160 and 190 GeV. The period of a SPS cycle is around 15 seconds, which includes the spill with a duration of 4.8 seconds. The full experimental setup of NA64-$\mu$ is detailed in~\cite{Gninenko:2014pea}. We highlight relevant aspects of the detector segment in~\figref{setupna64}. 

A calibrated muon beam with energy 150 GeV is injected into an active target with a length of around 20 cm (the length of the surrounding ECAL). The momentum and energy of the outgoing scattered muons are measured by a set of trackers. Ref.~\cite{Gninenko:2014pea} proposed to use a set of eight straw-tube chambers with a momentum resolution of $\sigma(p_\mu)/p_\mu = 3\%$ (for muon momentum $p_\mu = 100 \gev$) and 1 mm length resolution. Subsequent studies of the experimental setup~\cite{na64web} investigated alternative options, such as incorporating micromegas chambers or silicon based trackers, to further improve the momentum resolution\footnote{We thank Dipanwita Banerjee for providing insight into ongoing NA64-$\mu$ experimental studies and preparation.}. Photons and other secondary particles at large angles, generated during scattering events, are rejected by the electromagnetic calorimeter (ECAL) surrounding the target and two veto counters downstream. Other secondary particles at small angles are detected in the four hermetic hadronic calorimeter (HCAL) modules. 

\begin{figure}[!tbp]
   \centering
   \includegraphics[width=0.33\textwidth]{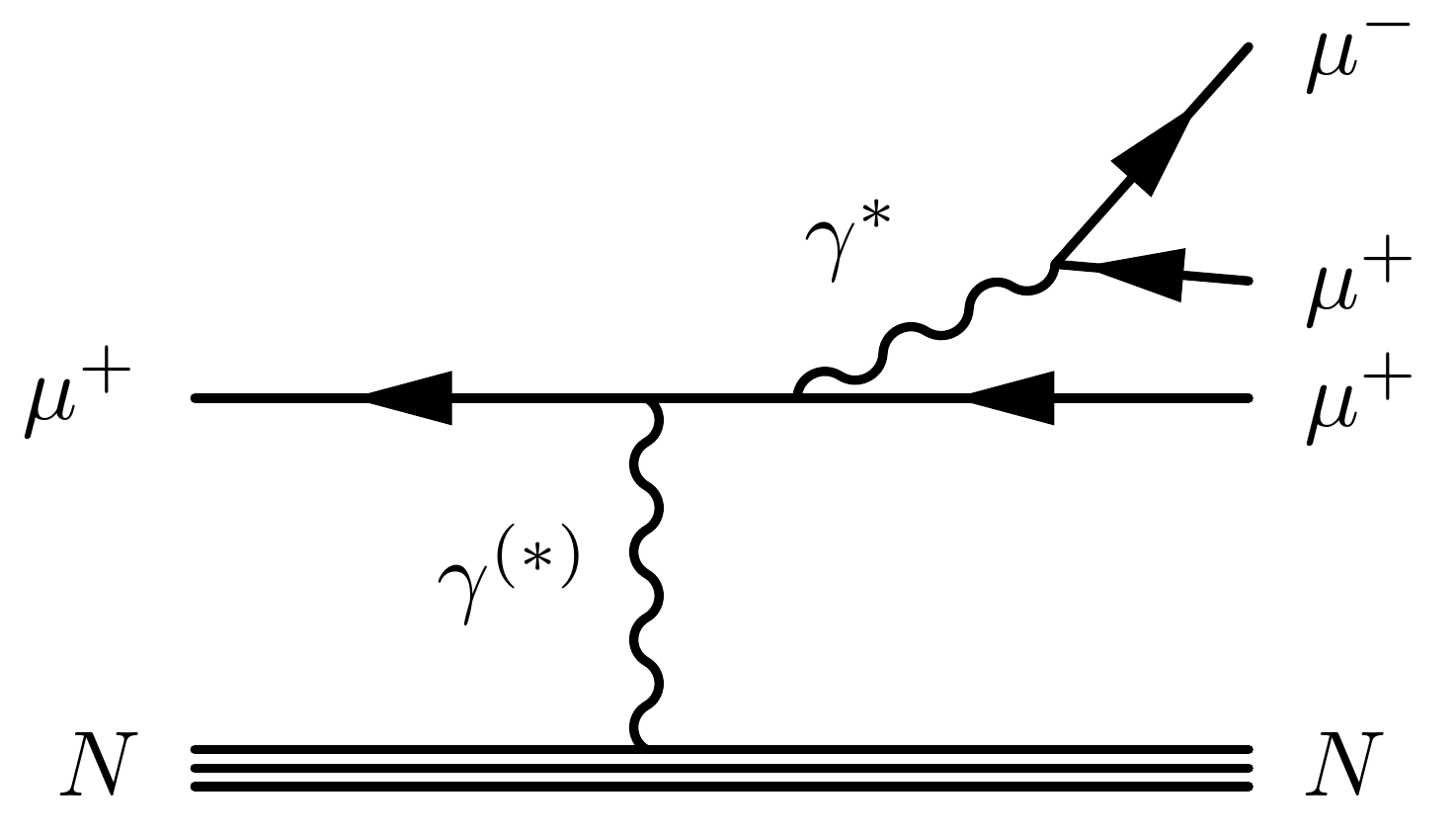}
   \quad
   \includegraphics[width=0.33\textwidth]{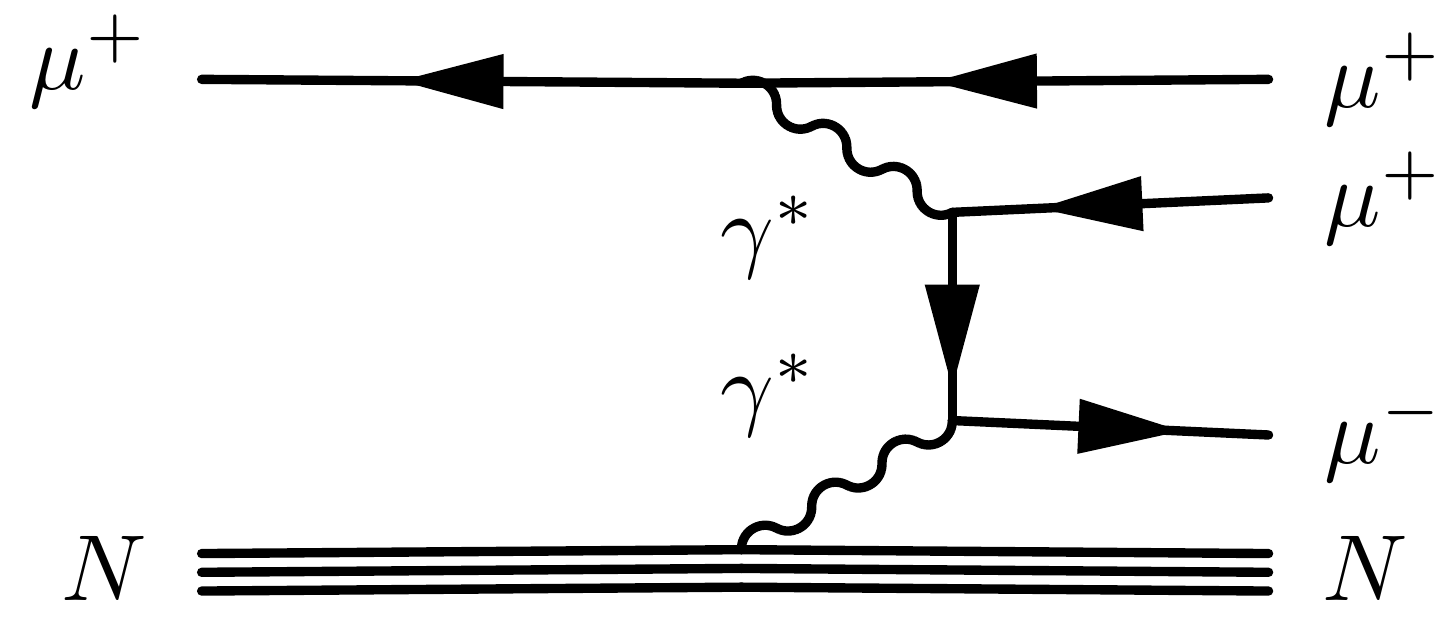} 
   \caption{Feynman diagrams contributing to the SM muon trident processes.}
   \label{fig:muontrident}
\end{figure}

A missing momentum or energy signal consists of a single scattered muon with energy $\leq 100 \gev$ with no accompanying energy deposition in the vetoes and a small amount of energy deposition in the ECAL and HCAL ($E_\text{ECAL} + E_\text{HCAL} \leq 12 \gev $). Ref.~\cite{Gninenko:2014pea} performed a detector Monte Carlo (MC) simulation given the above signal criteria. It was found that the dominant background arises from muon trident events, $\mu^+  N \to \mu^+  N (\mu^+ \mu^-)$ (see~\figref{muontrident}) at an expected level of $\lesssim 10^{-12}$ events per MOT. Other backgrounds are subdominant and are expected to be at the level of $\lesssim 10^{-13}$ events per MOT. 

The muon trident background can be challenging to eliminate when the momentum of the $\mu^+$ of the $\mu^+ \mu^-$ pair is much larger than the momentum of the $\mu^-$ in~\figref{muontrident}. It can fake a signal event if the soft $\mu^-$ is missed in the detector and the hard $\mu^+$ is so collinear with the scattered $\mu^+$ that they only produce a single track along the central region of the HCAL. Recently, Ref.~\cite{Kahn:2018cqs} proposed a  ``1 vs 2" method to further reject this background. The authors point out that the two collinear $\mu^+$s in the muon trident background are both minimum-ionizing particles (MIPs). Thus when passing through a layer of the HCAL, a background event would deposit roughly twice the energy and produce twice the number of photoelectrons as compared to a genuine signal event. Ref.~\cite{Kahn:2018cqs} suggested that the fake rate can be suppressed by a factor of $10^{-4}$ if the number of photoelectrons produced by a MIP is on the order of 100. 
While this ``1 vs 2'' method was originally proposed in the context of M$^3$, we suspect that a similar strategy can be incorporated at NA64-$\mu$, with a MIP producing $\simeq 150-200$ photoelectrons when passing through a single HCAL module~\cite{Gninenko:2014pea}. Therefore, using this method, a significant suppression of the muon trident background may be achievable at NA64-$\mu$. 

Given the considerations above, we suggest two background-free scenarios for a muon missing momentum search at NA64-$\mu$: (1) MOT $ =10^{12}$. This number is based on the background analysis in~\cite{Gninenko:2014pea}. Given the high intensity of the muon beam, it can be achieved in  a one-day run . (2) MOT $=10^{13}$. This would be a viable search option if  one adopts the ``1 vs 2" method and successfully reduces the muon trident background by at least an order of magnitude. This corresponds to about a nine-day run of the muon beam. For a given number of MOT, we estimate the 95\% confidence level (C.L.)~sensitivity by requiring $\geq3$ events, given our assumptions above. Longer run time with better background understanding can of course achieve even better sensitivity.

\subsection{Projections}

\begin{figure}[!t]
   \centering
   \includegraphics[width=0.45\textwidth]{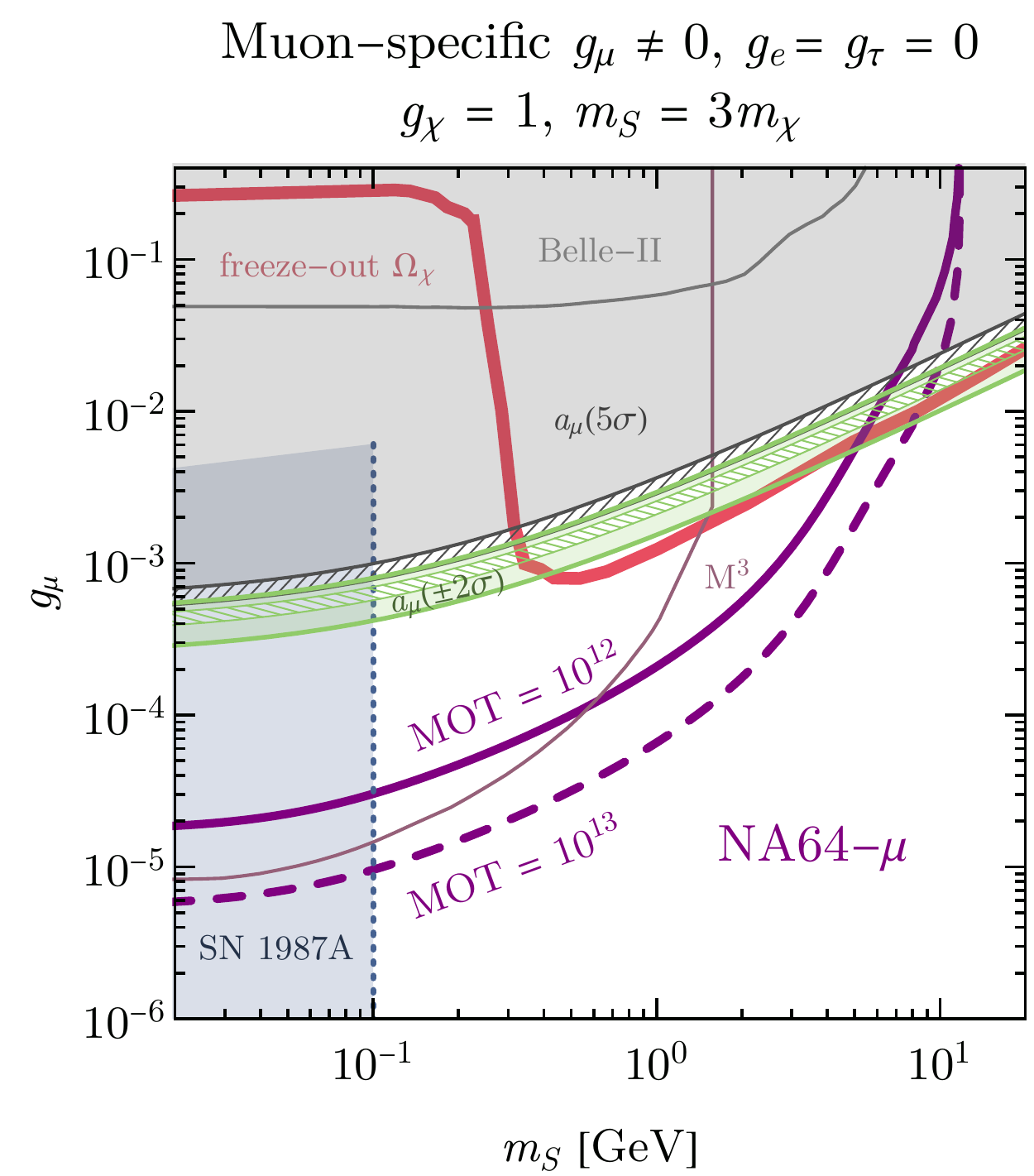} 
      \includegraphics[width=0.45\textwidth]{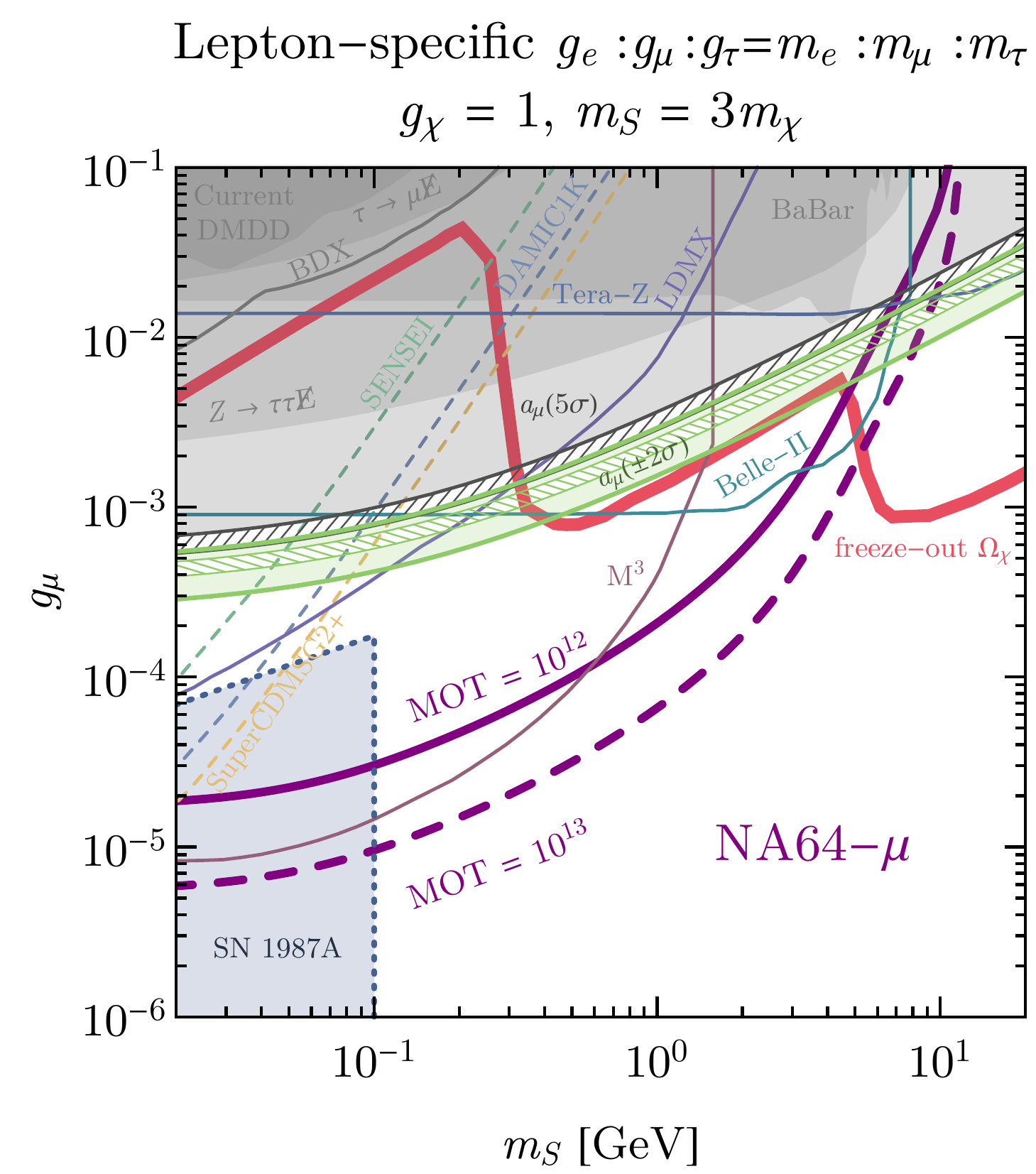} 
   \caption{95\% C.L. sensitivity projections for dark scalar searches at NA64-$\mu$ across the $g_\mu-m_S$ parameter space, along with several other experimental projections and existing constraints. The left and right panels correspond to the muon-specific ($g_\mu \neq 0$, $g_e = g_\tau =0$)  and lepton-specific ($g_e: g_\mu: g_\tau = m_e: m_\mu : m_\tau$) models, respectively. For the two models, we assume $m_\chi = m_S/3$ and $g_\chi =1$. The solid and dashed purple lines represent the expected NA64-$\mu$ sensitivity with $10^{12}$ and $10^{13}$ MOT, respectively. The red curve indicates the parameters required to reach the correct thermal freeze-out DM abundance. The green and gray regions represent the $2\sigma$-favored and $5\sigma$-excluded regions based on current $a_\mu$ measurements. The hashed regions indicate the projected sensitivity of future $a_\mu$ measurements, assuming the current central value stays the same while the experimental and theoretical uncertainties will be improved by a factor of 4 and 2, respectively. The blue region with dotted boundaries represents approximate exclusions from the cooling of SN 1987A. In both panels, we include the sensitivity projections for M$^3$~\cite{Kahn:2018cqs} and a Belle-II mono-photon search (with 50 fb$^{-1}$ data)~\cite{Dolan:2017osp}. For the lepton-specific model, we also include projections for LDMX (with 16 GeV electron beam)~\cite{Berlin:2018bsc}, BDX~\cite{Battaglieri:2017aum}, mono-photon searches at BaBar~\cite{Dolan:2017osp} and Tera-Z~\cite{Liu:2017zdh}, current LDM direct detection limits from XENON 10/100, DarkSide-50, and CDMS HVeV (collectively denoted as ``current DMDD''), expected DMDD sensitivities at SENSEI, DAMIC-1K, and SuperCDMS-G2+, as well as constraints from exotic $Z$ and $\tau$ decays. See text for further details.}
   \label{fig:money_plot}
\end{figure}

\begin{figure}[!t]
   \centering
   \includegraphics[width=0.45\textwidth]{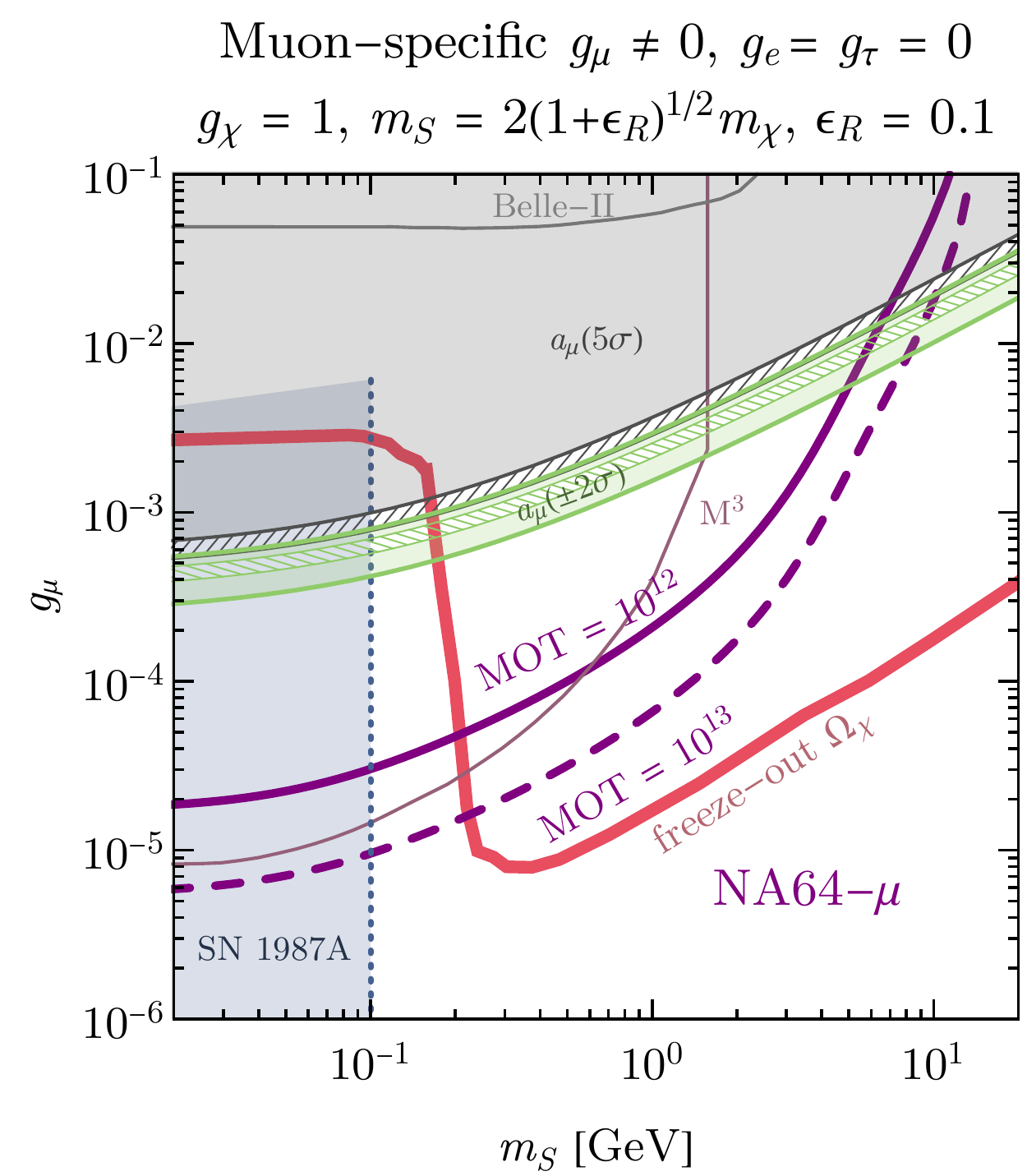} 
      \includegraphics[width=0.45\textwidth]{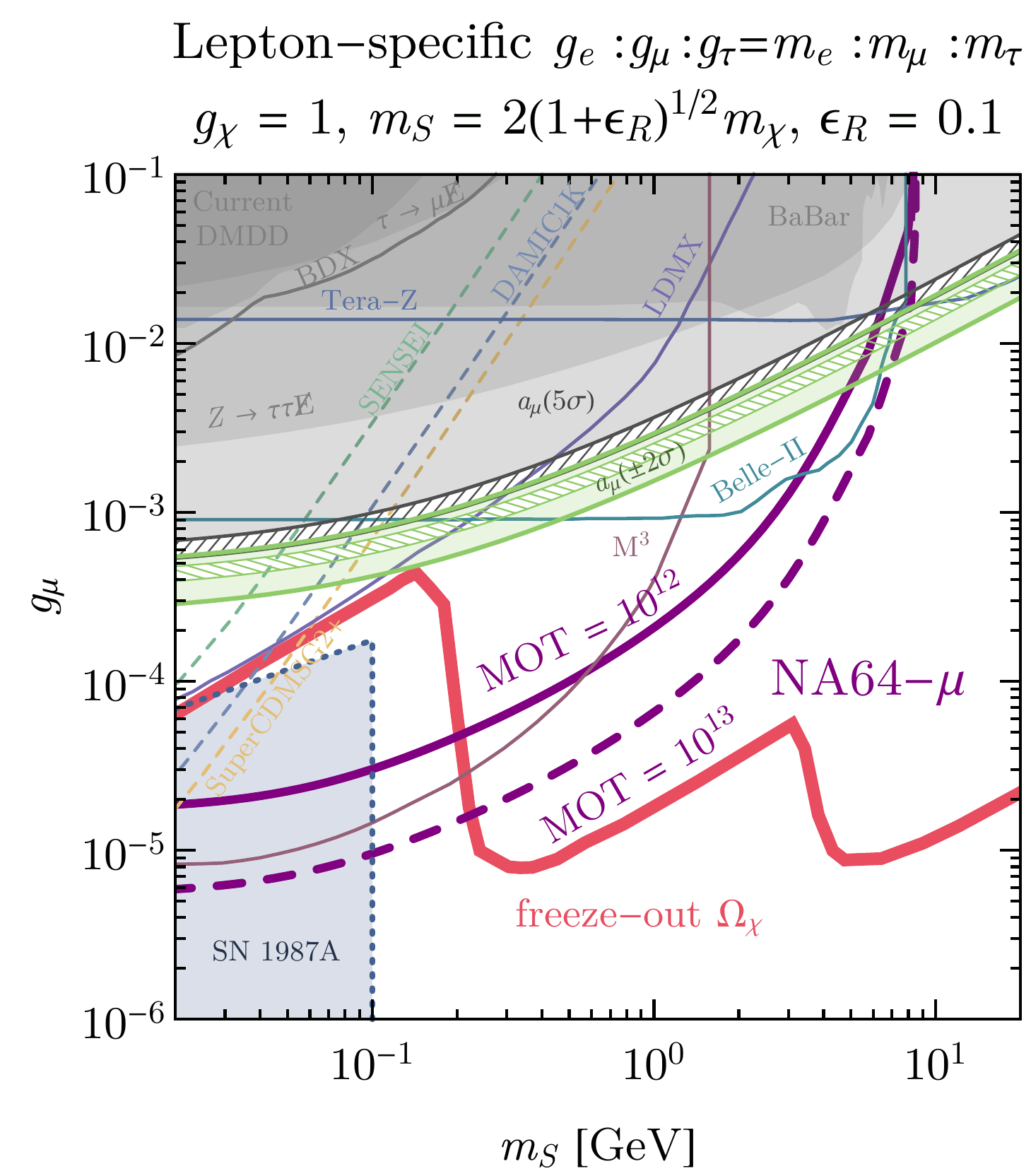}
   \caption{As in Fig.~\ref{fig:money_plot} but for $m_\chi = \f{m_S}{2 \sqrt{1+\epsilon_R}}$ with $\epsilon_R = 0.1$. The proximity of the resonance allows smaller values of the coupling to saturate the relic abundance while maintaining primarily invisible decays of the mediator across most of the parameter space. Once again, NA64-$\mu$ would be the most sensitive probe of the parameter space shown. }
   \label{fig:money_plot_2}
\end{figure}

The projected NA64-$\mu$ sensitivity to the models presented in Sec.~\ref{sec:models} is shown in Figs.~\ref{fig:money_plot} and~\ref{fig:money_plot_2}. The projections are shown both for $10^{12}$ (solid) and $10^{13}$ (dashed) MOT.  NA64-$\mu$ has the potential to probe the  $a_\mu$-favored region up to $m_S \sim 10$ GeV, due to the high beam energy. The sensitivity can be comparable to that of M$^3$ for $m_S < \OO(10 \mev)$, provided $10^{13}$ MOT can be achieved while running background-free. Figs.~\ref{fig:money_plot}-\ref{fig:money_plot_2} also show that NA64-$\mu$ can probe a considerable portion of the parameter space for which the relic abundance constraint for light leptophilic DM is satisfied via thermal freeze-out and without significant fine-tuning of $m_S/m_\chi$.

\section{Complementary probes}
\label{sec:complementarity}

Given the sensitivities achievable by NA64-$\mu$, it is natural to ask: what is the extent to which other current or planned experiments can explore the same models? We address this question here, showing that an NA64-$\mu$--type experiment is expected to probe significantly more of the model parameter space explaining the $(g-2)_{\mu}$ discrepancy and consistent with thermal relic dark matter than any other experiment. In muon-specific models, other relevant experimental probes include measurements of $(g-2)_\mu$ itself and supernova cooling bounds. For lepton-specific mediators with couplings to electrons and taus, there are several additional accelerator probes that can be relevant, including future searches at $B$-factories like Belle-II and proposed lepton collider ``$Z$-factories''. We discuss these various complementary probes in turn.

\subsection{Anomalous magnetic moments}

Current measurements of $a_{\mu}$ offer one of the most sensitive bounds on light scalars in muon-specific or lepton-specific models. In \figsref{money_plot}{money_plot_2}, the excluded parameter space for which $|a_\mu^{\rm NP} - \Delta a_{\mu}^{\rm central}| > 5 \sigma$ are shaded gray. Here, $a_\mu^{\rm NP}$ is the new physics contribution to $a_\mu$ from $S$ and $\Delta a_{\mu}^{\rm central}$ represents the central value of $\Delta a_\mu$ in~\eqref{g-2}. The excluded regions correspond to $g_\mu \gtrsim 10^{-3}$ at $m_S \sim \OO (10) \mev$ and $g_\mu \gtrsim \text{few} \times 10^{-2}$ at $m_S \sim 10 \gev$. 

Measurements of the anomalous magnetic moment of the electron, $a_e$, can also constrain the lepton specific model in principle. However the resulting sensitivity is significantly weaker due to the mass-proportional coupling hierarchy and is not competitive with other probes of the relevant parameter space in our models.

\subsection{Cooling of SN 1987A}
\label{sec:supernova}
A potentially important constraint on light scalars coupled to leptons arises from supernova cooling. A core-collapsed supernova behaves like a proto-neutron star. Its core region consists of highly-degenerate and relativistic electrons, near-degenerate and non-relativistic nucleons, and perhaps some amount of muons~\cite{Bollig:2017lki}. It cools mainly through neutrino diffusion. The measured SN 1987A neutrino burst flux agrees with SN model predictions~\cite{Raffelt:1996wa} and can be used to constrain dark scalars and leptophilic dark matter produced through its prompt decays\footnote{We thank Jae Hyeok Chang, Rouven Essig, and Harikrishnan Ramani for very helpful discussions about the dark scalar production and trapping mechanisms. We thank Jae Hyeok Chang for providing the trapping limit estimation.}. Similar to the SN studies for dark photons, if the DM-lepton coupling $g_\ell$ is too small, no significant dark scalar population can be produced in the SN and hence $S$ does not contribute significantly to cooling. On the other hand, if $g_\ell$ is too large, the DM produced through $S$ decays will be trapped inside the SN and, due to its frequent interactions with the SM plasma, again will not contribute to the cooling.  This implies a window of couplings (and masses) that can be constrained by supernovae.

Dark scalars with mass less than the plasma frequency of the photon, $\omega_p$ ($\simeq 20 \mev$ for SN 1987A), can be resonantly produced  through mixing with the longitudinal mode of the photon~\cite{Hardy:2016kme}. The resulting energy loss per unit mass is constrained by the Raffelt bound, $\epsilon_\text{Raffelt} = 10^{19}\, \text{erg}\, \text{g}^{-1} \text{s}^{-1}$. Requiring the energy loss to be smaller than this value yields an almost flat upper limit in the $m_S - g_{\mu}$ plane (the lower edge of the excluded region) around $g_e \approx 10^{-10}$  for $m_S \lesssim 20 \mev$~\cite{Knapen:2017xzo}. The bound on $g_e$ can be translated into $g_\mu = g_e m_\mu  /m_e \approx 10^{-8}$  for the lepton-specific model. If the dark scalar mass is much greater than the plasma frequency (e.g. $m_S \sim100 \mev$), resonant production is suppressed but the dark scalars can still be produced through continuum production. For example, the dark scalar can be produced through a Primakoff-like process $\gamma N \to  N S$~\cite{Batell:2017kty}. Ref.~\cite{Dolan:2017osp} considered this production mechanism for a pseudoscalar with photon coupling $(-g_{\gamma\gamma}/4) a F \tilde F$ and obtained an almost flat upper limit around $g_{\gamma\gamma} \approx 6\times 10^{-9} \gev^{-1}$ for $m_S\lesssim 100 \mev$. Ignoring the difference in CP quantum numbers and applying~\eqref{Lphoton}\footnote{Here we use the form factor $F_{1/2}$ with $q^2=0$. This approximation should be reasonable for the parameter space of interest, since supernovae are only expected to constrain $m_S$ significantly lighter than $2m_{\mu}$, and thus $|\rho| \ll 1$.}, the bound can be translated into  $g_\mu \approx \text{few}\times 10^{-7}$ for  the lepton-specific and the muon-specific models. Considering both the resonant and continuum production mechanisms, we conclude that an upper limit on $g_\mu$ exists around $\OO(10^{-8}-10^{-7})$ for scalar masses up to at least 100 MeV for both models. Of course, a much more careful analysis is needed to  combine all the possible production mechanisms and to properly account for plasma effects.

Once dark scalars are produced in the SN, they promptly decay into $\chi \bar{\chi}$ given our parameter choice $m_S > 2 m_\chi$ and $g_\ell \ll g_\chi =1$. Along their path of escape, DM interacts with the particles in the plasma, such as electrons, photons, and protons. Frequent interactions limit the DM outflow, and hence yield a lower limit (upper edge of the excluded region) on $g_\ell$ for scalars light enough to be produced significantly in the SN. For concreteness, we adopt the ``$\pi/2$-deflection criterion" proposed by~\cite{Chang:2018rso}: trapping is assumed to be sufficiently efficient provided that the expected accumulative deflection angle of DM particles is $\sim \pi/2$  along their path starting from the kinetic decoupling radius to the neutrino-gain radius ($\simeq 100 $ km).  Here we consider the trapping induced by $\chi e^- \to \chi e^-$ interactions to set a preliminary constraint. The corresponding trapping bound can be evaluated via Eqs.~(2.11-2.14) and Eqs.~(C1-C3) of~\cite{Chang:2018rso}.  Note that Eqs.~(C1-C3) need to be modified to account for the change from vector-mediated DM-nucleon interactions to scalar-mediated DM-electron interactions. The modification includes changing the mass, the number density, and the deflection angle per collision of nucleons to those of electrons. Also, one must include the correct electron distribution function in the phase-space integrals of Eqs. (C2-C3) to account for Fermi blocking. 

For the muon-specific model, $g_{\gamma\gamma}$-induced processes such as $\chi \gamma \to \chi \gamma$, $\chi N \to \chi N \gamma$, and $\chi e^- \to \chi e^- \gamma$ are likely to provide efficient trapping. In this case, one needs to account for different polarizations of the plasma photons. A detailed analysis of these trapping mechanisms is beyond the scope of this work. However, given that $g_{\gamma \gamma}$ is loop-- and $\alpha$--suppressed with respect to $g_e$, we expect that the upper edge of the excluded region for the muon-specific model in the $m_S-g_\mu$ parameter space will arise at larger couplings than in the lepton-specific model. To give a rough estimate of the trapping bound in this case, we consider only the $\chi \gamma \to  \chi \gamma$ process neglecting plasma effects, and adopt a simple mean-free-path criterion, requiring $(n_\gamma \sigma_{\chi \gamma \to \chi \gamma})^{-1} \lesssim r_\text{core}\approx 1\, \text{km}$. 

We show the resulting approximate SN 1987A bound for the lepton-specific model as a shaded blue region with dotted boundaries in~\figsref{money_plot}{money_plot_2}. The lower edge extends down to $g_\mu \sim \OO(10^{-8}-10^{-7})$ (below the plot range) and we cut the right edge off at $100$ MeV for a conservative estimate. The upper edge is around $g_\mu \simeq 10^{-4}$, increasing slightly for larger $m_S$ given that heavier dark matter ($m_\chi \propto m_S$) is more difficult to trap. A similar exclusion region is shown for the muon-specific model. Here the upper limit is at $\OO(10^{-3})$. We do not include a dotted boundary line here, given that our estimate excludes potentially important plasma effects and other trapping processes. We again emphasize that our limits are preliminary, and defer a more comprehensive analysis for the two models to~\cite{inprogress}. Nevertheless, the general message is clear: for light dark scalars with mass up to $\sim 100$ MeV, SN probes are complementary to muon missing momentum searches.

\subsection{Dark Matter Direct detection (DMDD)}

\begin{figure}[!t]
\centering
\includegraphics[width=.4\textwidth]{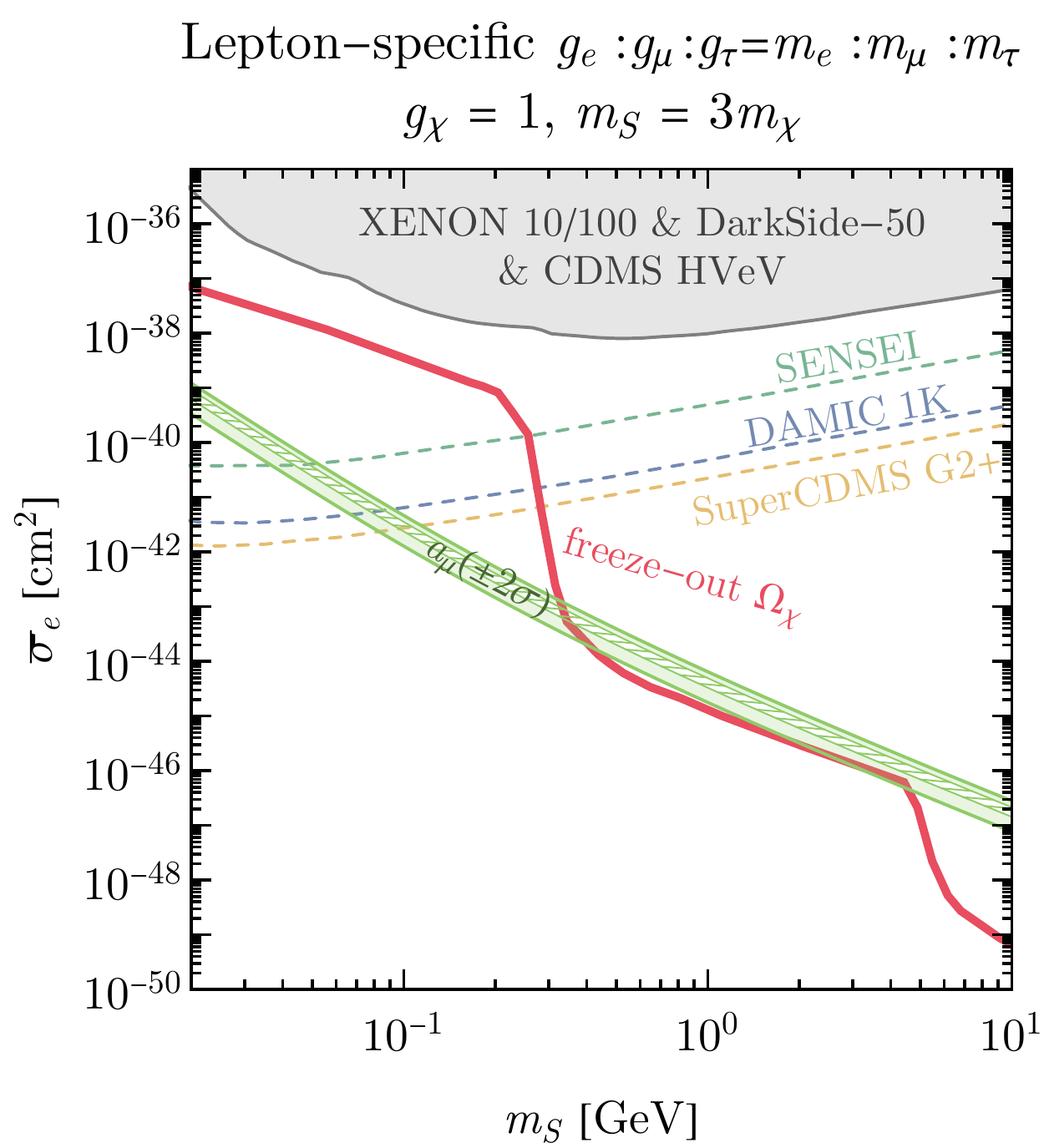} \quad \includegraphics[width=.4\textwidth]{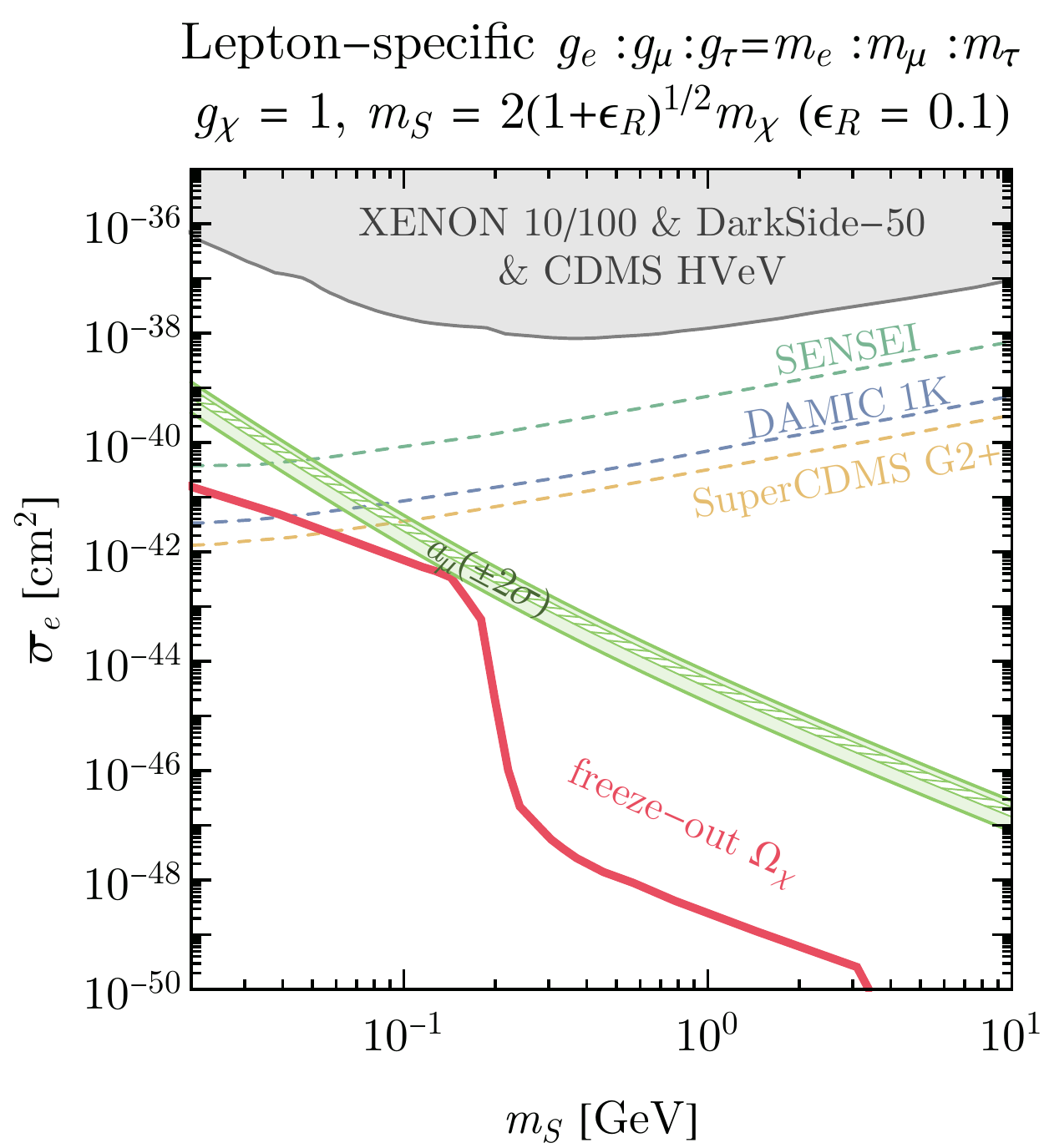}
\caption{\label{fig:direct_detection} The predicted dark matter--electron scattering cross-section, $\overline{\sigma}_e$, for a leptophilic scalar mediator with $g_{\chi}=1$,  and $m_\chi=m_S/3$ (left) and $m_S/(2 \sqrt{1+\epsilon_R})$ with $\epsilon_R=0.1$ (right). Also shown are constraints and projections from present and future direct detection experiments. Regions above the experimental sensitivity curves are expected to be probed. Values of $\overline{\sigma}_e$ corresponding to the couplings $g_{e}$ required to saturate the relic abundance in each case are given by the red contours. The current (future) $a_{\mu}$-favored band is also indicated by the shaded (hashed) green regions.}
\end{figure}

In the leptophilic scenario described above, the scalar mediates scattering between $\chi$ and electrons.\footnote{$g_{\gamma\gamma}$ -induced scattering processes, such as $\chi N \to \chi N \gamma$ and $\chi e^- \to \chi e^- \gamma$, are negligible due the $\alpha-$, loop-, and phase-space-suppression.} This can lead to a signal in direct detection experiments sensitive to electron recoils. To determine the sensitivity of such experiments to the models of interest, we follow~\cite{Essig:2011nj, Essig:2015cda,  Battaglieri:2017aum} and define a reference $\chi-e^-$ scattering cross-section, $\overline{\sigma}_e$ with momentum exchange set to $q = \alpha m_e $, and a form factor, $F_{\rm DM}(q)$ to account for the $q$-dependence. For the mediator masses of interest, $m_S \gg \alpha m_e$ (the characteristic momentum scale for atomic processes), and so $F_{\rm DM}(q) \simeq 1$. The reference cross-section in the non-relativistic limit is  
\begin{equation}
\overline{\sigma}_e \simeq \frac{\mu_{\chi e}^2 g_{\chi}^2 g_e^2}{\pi m_{S}^4},
\end{equation}
where $\mu_{\chi e}$ is the dark matter -- electron reduced mass. In the parameter space of interest, $\mu_{\chi e}\approx m_e$ and the direct detection limits are generally weakly sensitive to the DM--scalar mass ratio. Nevertheless, we show the predicted value of $\overline{\sigma}_e$ for the leptophilic DM parameter space saturating the observed dark matter relic abundance with $g_{\chi}=1$ and $m_{\chi}=m_S/3$, $m_{\chi} = m_S/(2 \sqrt{1+\epsilon_R})$ with $\epsilon_R = 0.1$  in the left- and right-hand panels of Fig.~\ref{fig:direct_detection}, respectively. We also show the corresponding $\overline{\sigma}_e$ values for the $a_{\mu}$-favored regions.

The XENON10~\cite{Essig:2012yx,Essig:2017kqs, Angle:2011th}, XENON100~\cite{Essig:2017kqs, Aprile:2016wwo}, DarkSide-50~\cite{Agnes:2018oej}, and CDMS HVeV~\cite{Agnese:2018col} experiments already set limits on $\overline{\sigma}_e$ in the relevant mass range.\footnote{Constraints on DM-electron scatterings from the excess of earth heat flux~\cite{Chauhan:2016joa} are much weaker than current DMDD constraints for the range of parameters shown in~\figref{money_plot},~\ref{fig:money_plot_2}, and~\ref{fig:direct_detection}. Hence we do not show them.} Excluded values are shaded in Fig.~\ref{fig:direct_detection}. These experiments are not currently sensitive to the leptophilic parameter space saturating the relic abundance or explaining the $(g-2)_{\mu}$ discrepancy for our choices of parameters. There are, however, many proposed and planned direct detection experiments that are expected to substantially improve the reach in $\overline{\sigma}_e$ for light dark matter (see~\cite{Battaglieri:2017aum} for an overview of these efforts). In Fig.~\ref{fig:direct_detection}, we show projections for three such experiments: SENSEI with an $\mathcal{O}(100\,{\rm gram})$ detector, DAMIC-1K, and SuperCDMS; all projections are taken from~\cite{Battaglieri:2017aum}. There are other experimental proposals potentially sensitive to the relevant parameter space, but the projections shown demonstrate the range of cross-sections expected to be reached by next-generation experiments. 

From Fig.~\ref{fig:direct_detection}, we see that next-generation direct detection experiments can probe the leptophilic DM parameter space with a lepton-specific scalar mediator consistent with the measured value of $a_{\mu}$ for $m_S \lesssim 100$ MeV (assuming $g_e/g_{\mu} \sim m_e/m_{\mu} $). For the benchmark case $m_{\chi}=m_S/3$, $g_{\chi}=1$, this region is inconsistent with the observed relic abundance. However, this is not necessarily the case for $m_{\chi} \sim m_S/2$ (near the $s-$channel resonance), in which both future direct detection experiments and muon beam experiments such as NA64-$\mu$  can be sensitive to the $a_{\mu}$-favored regions, as illustrated by the results for the near-resonant benchmark point in \figref{direct_detection}. This opens up the exciting possibility of detecting both dark matter and its mediator to the SM within the next generation of experiments, as well as resolving the $(g-2)_{\mu}$ puzzle. Of course in the muon-specific case where the scalar does not couple to electrons, direct detection experiments will not be sensitive to dark matter interactions with the SM, and instead a muon missing momentum experiment will be crucial for the discovery of the dark sector.

\subsection{Collider experiments}

\begin{figure}[!t]
    \centering
\begin{subfigure}[b]{0.28\textwidth}
        \includegraphics[width=\textwidth]{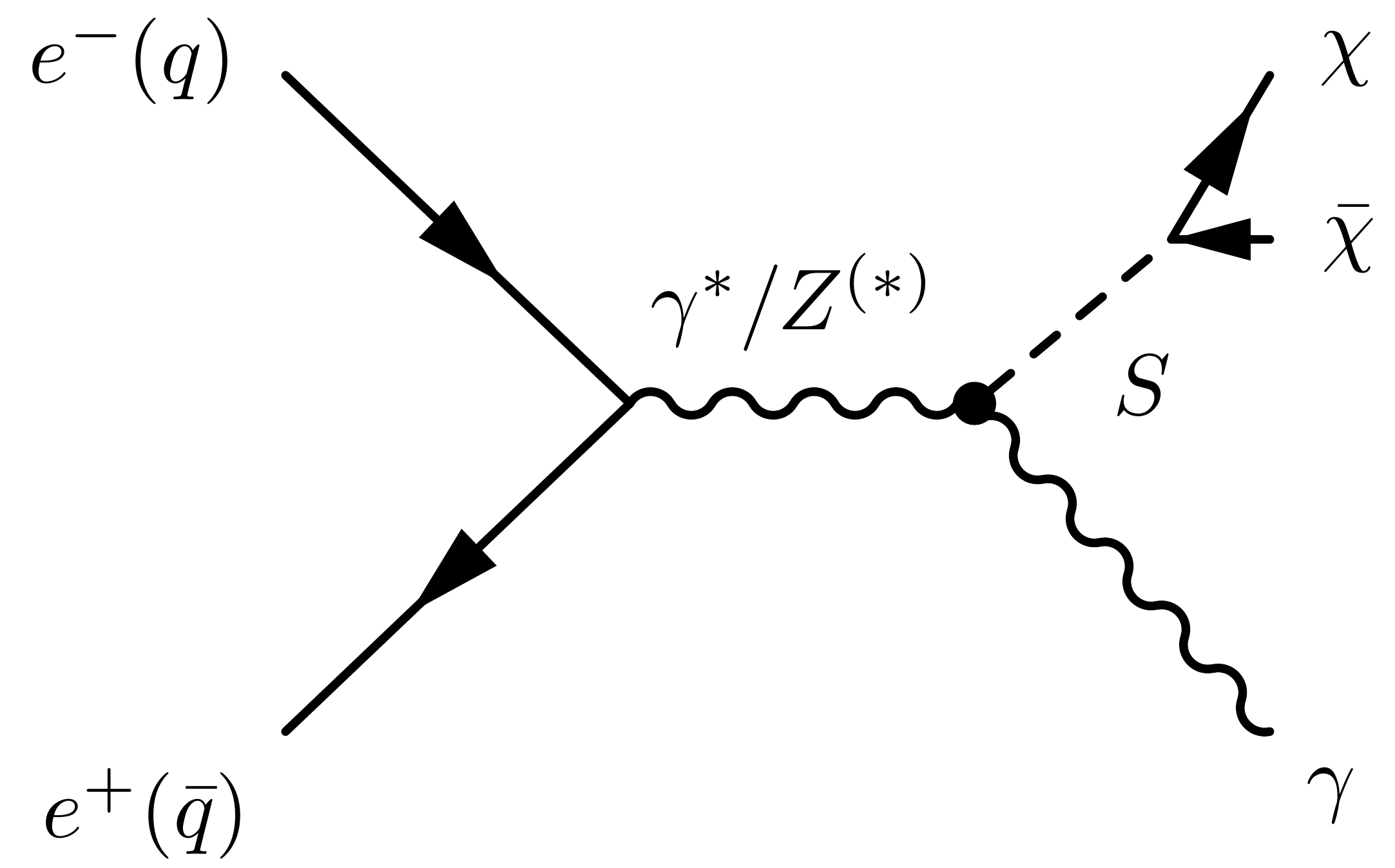}
        \caption{}
        \label{fig:collider2}
 \end{subfigure}
                 \quad\quad\quad
                 \begin{subfigure}[b]{0.3\textwidth}
        \includegraphics[width=\textwidth]{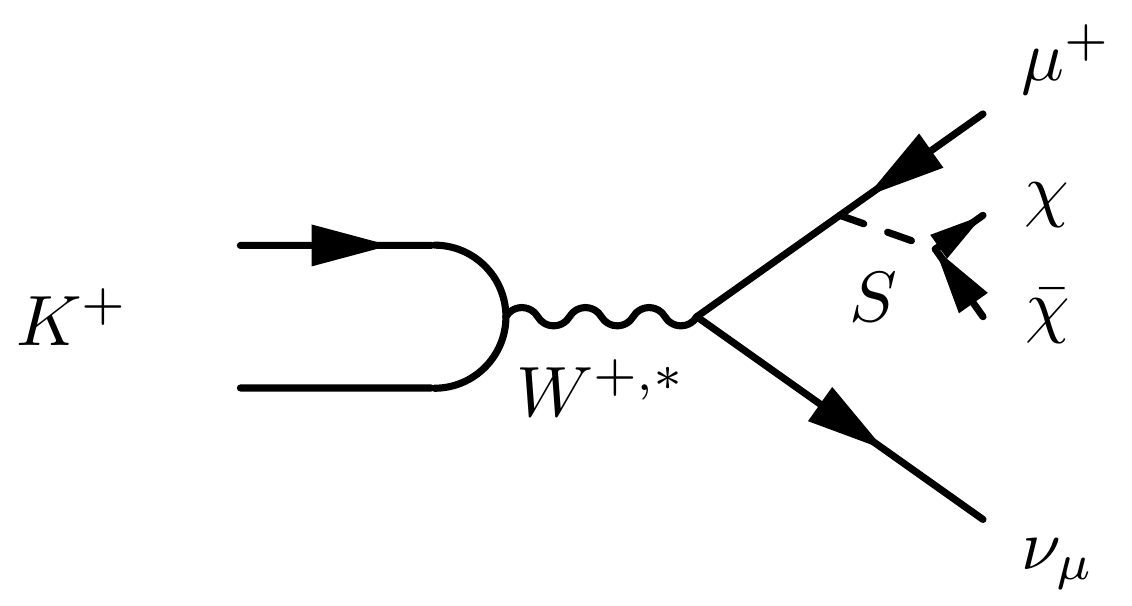}
        \caption{}
        \label{fig:sk}
\end{subfigure}
                \quad\quad\quad\quad
 \begin{subfigure}[b]{0.27\textwidth}
        \includegraphics[width=\textwidth]{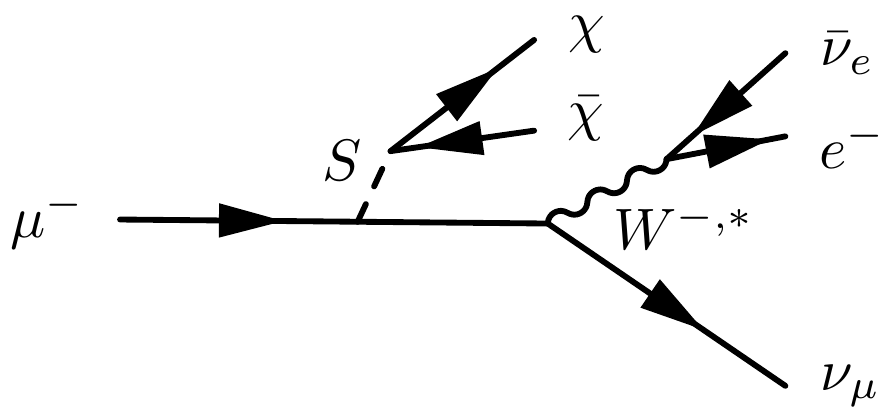}
        \caption{}
        \label{fig:slla}
\end{subfigure}
    \quad\quad
     \renewcommand{\thesubfigure}{d-1} 
\begin{subfigure}[b]{0.28\textwidth}
        \includegraphics[width=\textwidth]{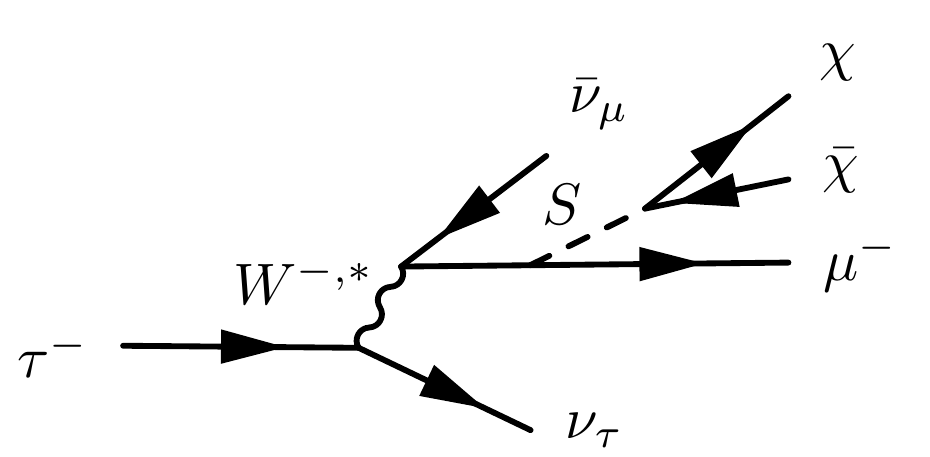}
        \caption{}
       \label{fig:sllb}
 \end{subfigure}
       \quad\quad 
       \renewcommand{\thesubfigure}{d-2} 
\begin{subfigure}[b]{0.27\textwidth}
        \includegraphics[width=\textwidth]{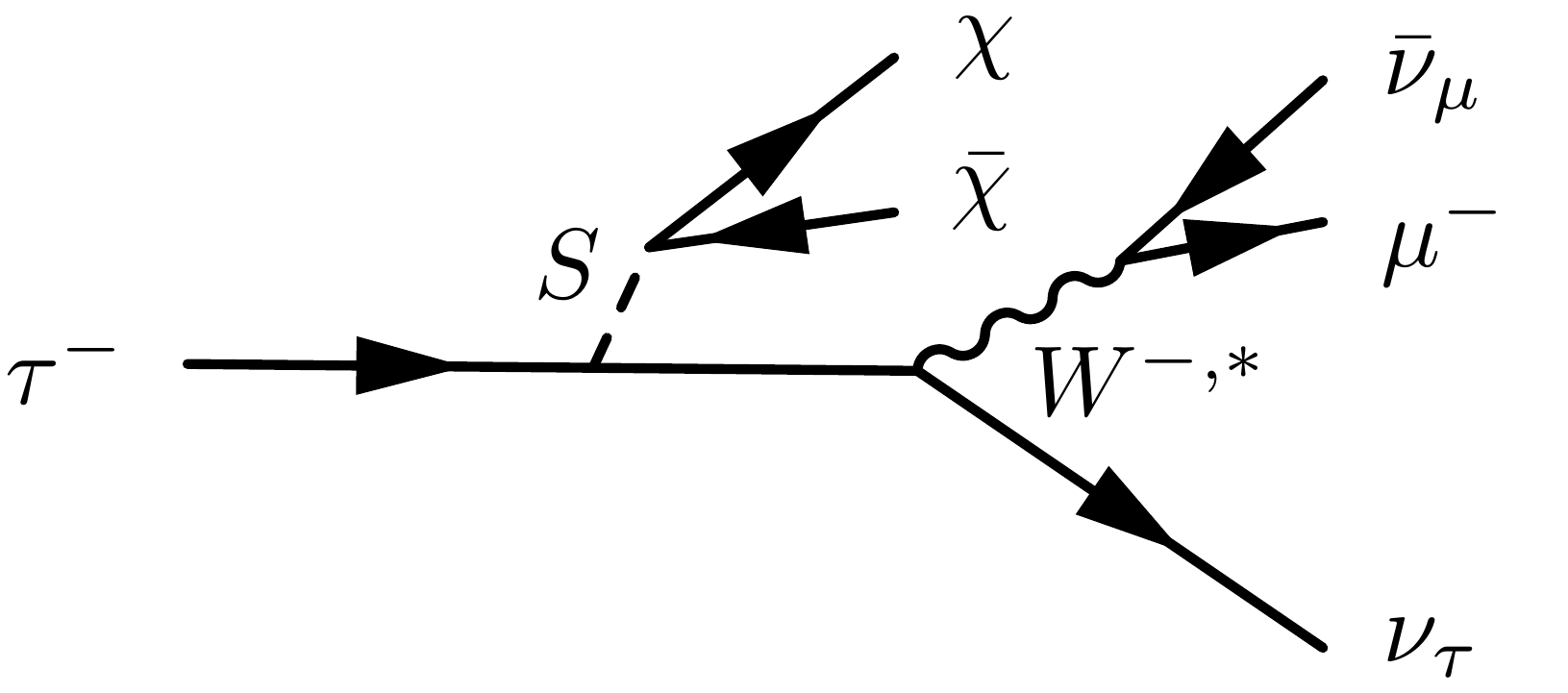}
        \caption{}
        \label{fig:sllc}
\end{subfigure}
               \quad
\caption{Diagrams sensitive to $g_\mu$ or the $g_\mu$--induced $g_{\gamma\gamma}$ coupling in the muon-specific model and contributing to (a) $e^+ e^- (q \bar{q}) \to \gamma S$, (b) $K^+ \to \mu^+ \nu_{\mu} S$, (c) $\mu^- \to e^- \nu_\mu \bar{\nu}_e S$ , and (d-1,d-2) $\tau^- \to \mu^- \nu_\tau \bar{\nu}_\mu S$, with $S$ subsequently decaying to $\chi\bar{\chi}$.}
    \label{fig:muondecay}
\end{figure}

There are several collider probes of light leptophilic invisibly-decaying scalars. $e^+ e^-$ experiments can be sensitive to $S$ through the process $e^+ e^- \rightarrow \gamma S(\rightarrow \overline{\chi}\chi)$ in mono-photon searches. In the muon-specific model, $S$ couples to photons through a muon-loop (cf.~\eqref{Lphoton}), giving rise to a signal in this channel through the diagram in Fig.~\ref{fig:collider2}. In the lepton-specific scenario, the $S\gamma \gamma$ coupling receives contributions from all three lepton flavors. The coupling of $S$ to electrons results in additional $t/u$-channel diagram contributions to $e^+ e^- \rightarrow \gamma S(\rightarrow \overline{\chi}\chi)$. Across the parameter space of interest, however, we find that the $S\gamma\gamma$ constribution represented by Fig.~\ref{fig:collider2} dominates the production rate, even in the lepton-specific model. To place limits on the $m_S-g_\mu$ parameter space, we therefore utilize the 90\% C.L. upper bounds and projections from BaBar and Belle-II in Ref.~\cite{Dolan:2017osp}. In particular, we compute the $e^+ e^- \rightarrow \gamma S(\rightarrow \overline{\chi}\chi)$ production cross-section, utilizing the form factor in Appendix~\ref{sec:app} with $q^2 = s$, the center-of-mass energy for BaBar or Belle-II, and compare to the cross-section for axion-like pseudoscalars in Ref.~\cite{Dolan:2017osp}. We find that the bounds in our scalar case are weakened by a factor of $\sim \sqrt{1.8}$ relative to the pseudoscalar bounds in Ref.~\cite{Dolan:2017osp} across the angular regions accessible by BaBar. Assuming a similar angular acceptance at Belle-II, the same factor of $\sqrt{1.8}$ applies\footnote{If, for example, all angles were accepted, the factor of $\sqrt{1.8}$ would reduce to a factor of $\sqrt{1.5}$ and not significantly impact our results.}. We therefore interpret the BaBar bounds and Belle-II projections on $g_{a\gamma\gamma}$ from Ref.~\cite{Dolan:2017osp} as bounds on $g_{\gamma \gamma}$, including this factor of $\sqrt{1.8}$. In addition, we scale up the coupling by $\sqrt{1.96/1.64}$ to estimate 95\% C.L. limit based on the 90\% C.L. limit.

The resulting limits and projections are shown in Figs.~\ref{fig:money_plot}-\ref{fig:money_plot_2}. We find that mono-photon search at BaBar is not sensitive to any of the parameter space not already excluded by $a_{\mu}$ measurements in either the muon- or lepton-specific models. In the lepton-specific case, mono-photon search at Belle-II with 50\, ab$^{-1}$ will be able to probe the $a_\mu$--favored regions of the model with $m_S\sim 100-1000$ MeV. Note that this measurment is an \emph{indirect} probe of the couplings to leptons, and an experiment like NA64-$\mu$ would still be required to conclusively determine whether or not $S$ couples to muons. 

Other collider searches yield less sensitivity. Experiments such as BaBar and LHC$b$ can probe the $S$-muon coupling through the process $e^+ e^- (q \bar{q}) \rightarrow \mu^+ \mu^- S$. BaBar has searched for $e^+ e^- \to \mu^+ \mu^- S(\rightarrow \mu^+ \mu^-)$~\cite{TheBABAR:2016rlg} and placed constraints on visibly-decaying mediators. However, we are not aware of analogous searches for $\mu^+ \mu^- + \slashed{E}$. This likely has to do with the fact that, in the visible case, the $\mu^+ \mu^-$ final state can be reconstructed as a narrow resonance that provides powerful discrimination from the background, whereas in the invisible case, no such discrimination is possible. The same reasoning likely applies to LHCb as well. KLOE~\cite{Babusci:2015zda} has performed a   $\mu^+ \mu^- + \slashed{E}$ search for $e^+ e^- \to A' S \to (\mu^+ \mu^-) (\chi \bar{\chi})$ where a bump hunt in $m_{\mu^+\mu^-}$ can be performed. Given that $S$ decays primarily invisibly in the parameter space we are interested in, we do not expect a relevant constraint can be inferred from the search. 

$S$ can also be radiated off of muons produced in kaon decays at accelerator experiments, providing another direct probe of the coupling $g_\mu$. The most relevant bound comes from searches for $K^+ \to \mu^+ \nu_\mu + \slashed{E}$, with the corresponding production mechanism shown in Fig.~\ref{fig:sk}. The E949 experiment at Brookhaven limits BR$(K^+ \to \mu^+ \nu_\mu + \slashed{E}) \lesssim (2-3)\times 10^{-6}$, depending on the kinematics of the BSM contribution~\cite{Artamonov:2016wby}. In Ref.~\cite{Chen:2015vqy}, the branching ratio for $K^+ \to \mu^+ \nu_{\mu} S$ was calculated to be $\lesssim 10^{-8}$ for couplings $g_{\mu}$ large enough to explain the $(g-2)_{\mu}$ discrepancy, with smaller $g_{\mu}$ resulting in even smaller branching ratios. We thus conclude that these kaon decay experiments are not sensitive to the parameter space of interest. 

Similarly, pion decays could in principle provide another probe of $S$ in the low mass range. Current limits on pion branching ratios are expected to yield a weaker bound than that from kaons: the relevant kaon constraint comes from a dedicated search for $K^+ \to \mu^+ \nu_{\mu}+ \slashed{E}$~\cite{Artamonov:2016wby} and exploits the fact that the muon in such decays will have a smaller momentum than those in e.g.~$K^+ \to \mu^+ \nu_{\mu} $. This allows for a significant suppression of the SM contributions. We are not aware of an analogous search in the pion case, and so we do not include a corresponding pion decay bound in Table 1. On-going and future experiments, such as SeaQuest, can provide sensitivity to visibly-decaying light scalars with couplings to muons produced in pion decays, as discussed in Ref.~\cite{Berlin:2018pwi}. However, if $S$ decays primarily invisibly, as we have assumed across the parameter space of interest, then the relevant final states involve only prompt leptons and missing energy, and are more difficult to probe. We are not aware of any planned searches for rare pion decays that would provide additional sensitivity to the allowed regions in Figs.~\ref{fig:money_plot} and~\ref{fig:money_plot_2}. In any case, pion decays are only relevant in the low-mass regions of parameter space which are already constrained by SN1987A. 

\begin{table}[t]
   \centering
      \begin{tabular}{@{} lr @{}} 
      \hline
            Decay process & 95\% C.L. upper limit on BR\\
      \hline
       $K^+ \to \mu^+ \nu_\mu S $ & $ (2-3)\times 10^{-6}$\\
       $Z \to \gamma S$ & $ 1.1\times 10^{-6}$\\
       $Z \to \tau^+ \tau^- S$ & $ 5.1\times 10^{-3}$\\
          $\mu^- \to e^- \nu_\mu \bar{\nu}_e S$ & $ 2.0\times 10^{-6}$ \\
            $\tau^- \to \mu^- \nu_\tau \bar{\nu}_\mu S$ &  $ 8\times 10^{-4}$ \\
      \hline
   \end{tabular}
   \caption{Summary of rare decays that can probe $g_\mu$ in the muon-specific scenario. The 95\% C.L. upper limits on the branching ratios are estimated based on~\cite{Artamonov:2016wby} for $K$ decays and data in~\cite{Olive:2016xmw} for $Z$, $\mu$, and $\tau$ decays. Here $S$ is assumed to decay invisibly.}
   \label{tab:pdgemu}
\end{table}

\subsection{$Z$ decays}

At higher energies, precision measurements of $Z$ boson properties can also place constraints on the $g_\mu-m_S$ plane. In the muon-specific model, the corresponding constraints are weak. The best sensitivity arises in regions of the parameter space where BR$(S \to \mu^+ \mu^-)$ is non-negligible, so that one can obtain constraints from $Z \to \mu^+ \mu^- S (\to \mu^+ \mu^-)$. This process contributes to the $4\mu$ decay of the $Z$, which has a measured branching ratio of ${\rm BR}(Z \to \mu^+ \mu^- \mu^+ \mu^-)=(3.5 \pm 0.4) \times 10^{-6}$~\cite{Olive:2016xmw} . However, we find that deviations at the level of these uncertainties still only arise for large couplings and masses that are already excluded by $(g-2)_\mu$ measurements. There can also be a bound from BR$(Z \to \gamma S)$, reflected in Tab.~\ref{tab:pdgemu}. The current limit is set by an L3 search at LEP~\cite{Acciarri:1997im}, BR$(Z\to \gamma S) < 1.1\times 10^{-6}$, where the energy of the photon is required to be greater than $\sim 30$ GeV. The decay $Z \to \gamma S$ occurs through the $SZ\gamma$ coupling induced through a muon-loop (analogous to the $S\gamma \gamma$ coupling), followed by $S \to \chi \bar{\chi}$. The expressions for the loop-induced $Z\to \gamma S$  decay width can be found in Refs.~\cite{Gunion:1989we, Djouadi:2005gi}.The corresponding constraint again turns out to be very weak due to the Yukawa and loop suppression.

In the lepton-specific case, the couplings to $\tau$'s can yield stronger sensitivity. In particular, the process $Z \to \tau^+ \tau^- S (\to \chi \chi)$ can contribute to the measured $Z \to \tau^+ \tau^-$ width, assuming that the kinematics associated with radiating the scalar $S$ still allow events to pass the signal acceptance criteria. The decay width $\Gamma(Z \to \tau^+ \tau^-)$ is measured to be $\Gamma(Z \to \tau^+ \tau^-) = 84.08 \pm 0.22 \;\; {\rm MeV}$~\cite{Olive:2016xmw}. To obtain an approximate 95\% C.L. limit, we simply require that the width $\Gamma(Z \to \tau^+ \tau^- S, S \to \chi \chi)$ be less than 1.96 times the uncertainty of the $\Gamma(Z \to \tau^+ \tau^-)$ measurement. 
The resulting approximate limit is shown in Fig.~\ref{fig:money_plot}. Despite the larger coupling to $\tau$'s, this constraint is still not strong enough to exclude the $a_\mu$ favored region. The bound cuts off around $\sim 10$ GeV, since for larger masses the branching fraction of $S \to \chi \chi$ becomes significantly less than one for $g_\mu \gtrsim 0.1$ due to the large $g_{\tau}$. Note that the corresponding constraints from $Z \to e^+ e^-$ or $Z \to \mu^+ \mu^-$ are much weaker given the coupling hierarchy. The lepton-specific case also induces $Z \to \gamma S$ decays, which, when followed by $S \to \chi \chi$, can be bounded as in Tab.~\ref{tab:pdgemu}. Again, however, only large couplings are excluded,  $g_\mu \gtrsim 0.7$.

Future lepton colliders such as the FCC-ee~\cite{dEnterria:2016sca} or CEPC~\cite{CEPC-SPPCStudyGroup:2015csa} could place stronger bounds on  $Z\to S\gamma$. Both proposals plan to run at the $Z$-pole and can potentially produce $10^9$ --$10^{12}$ $Z$ bosons.  We infer a sensitivity projection for a Tera-Z factory (corresponding to $~10^{12}~Z$ bosons) using the results of Fig. 9(C) in Ref.~\cite{Liu:2017zdh}, which shows the reach for $Z$ boson decays to a photon and pseudo-scalar, followed by the invisible decay of the pseudo-scalar. The resulting projection is shown in Figs.~\ref{fig:money_plot}- \ref{fig:money_plot_2}.  Although the expected sensitivity mostly lies within the $5 \sigma$ $(g-2)_\mu$ exclusion region, it is important to notice that a Tera-Z factory could potentially probe the $a_\mu$--favored region for $m_S$ greater than 10 GeV. The reach cuts off around $m_S\sim 80$ GeV due to the phase space suppression. Other Tera-Z searches, such as $Z\to \tau^+ \tau^- + \slashed{E}$, could also in principle be relevant. But the corresponding sensitivity can be largely restricted by experimental systematic uncertainties and we leave a more detailed study in future. Also, above the $Z$ pole, other high-energy collider searches might apply, however, since we are focused on light dark sectors below the electroweak scale, we defer their consideration to future study.

\subsection{$\mu$ and $\tau$ decays}

Given non-zero $g_\mu$, the scalar $S$ can be radiated off of a muon in  $\mu^- \to e^- \nu_\mu \bar{\nu}_e $ and  $\tau^- \to \mu^- \nu_\tau \bar{\nu}_\mu $ decays. The corresponding diagrams are shown in Figs.~\ref{fig:slla} and~\ref{fig:sllb}. The coupling $g_\mu$ can thus be constrained by measurements of the 
 $\mu$ and $\tau$ decay widths. We have computed $\Gamma(\mu^- \to e^- \nu_\mu \bar{\nu}_e S)$ and $\Gamma(\tau^- \to \mu^- \nu_\tau \bar{\nu}_\mu S)$ for $m_S \leq m_\mu-m_e$ and $m_\tau-m_\mu$, respectively, in \texttt{Madgraph5 2.5.1}~\cite{Alwall:2014hca} and compared them with the upper limits on the branching fractions listed  in~\tabref{pdgemu}. The muon decay bound was derived by requiring $\Gamma(\mu^- \to e^- \nu_\mu \bar{\nu}_e S)$ to be smaller than the uncertainties in the measured muon total width reported in \cite{Olive:2016xmw}. We find that, at 95\% C.L., the measured muon lifetime generally constrains couplings $g_{\mu}$ larger than 1. The $\tau$ decay bound in~\tabref{pdgemu} was derived by requiring $BR(\tau^- \to \mu^- \nu_\tau \bar{\nu}_\mu S)$ to be smaller than the uncertainties on the measured $\tau^- \to \mu^- \nu_\tau \bar{\nu}_\mu$ branching ratio tabulated in~\cite{Olive:2016xmw}. This yields somewhat stronger bounds on the coupling, constraining $g_\mu \lesssim 0.5$ at 95\% C.L. for $m_S \approx 20 \mev$. The bound weakens as $m_S$ increases. We conclude that these bounds do not constrain the parameter space of interest in the muon-specific model.

Similarly to the muon-specific case, $S$ can be produced in $\mu$ and $\tau$ decays in the lepton-specific scenario. In addition to the diagrams in~\figref{slla} and \figref{sllb}, $S$ can also be radiated off of taus in the initial state, as in \figref{sllc}. \footnote{There is also a similar diagram for $S$ radiated off from $\tau$ or $e$ in $\tau^- \to e^- \nu_\mu \bar{\nu}_e $ and $\tau^- \to e^- \nu_\tau \bar{\nu}_e$. Given the similar branching ratio of $\text{BR}(\tau \to e+ \slashed{E}$) and $\text{BR}(\tau \to \mu + \slashed{E}$) and $g_e \ll g_\mu$ in the lepton-specific model, the latter search yields a stronger bound.} While $\mu^- \to e^- \nu_\mu \bar{\nu}_e S$ still only constrains $g_\mu \gtrsim \OO (1)$, the decay $\tau^- \to \mu^- \nu_\tau \bar{\nu}_\mu S$ restricts $g_{\mu} \lesssim 0.02$ for $m_S \approx 20 \mev$ at 95\% C.L.. The bound is stronger than in the muon-specific case due to the $m_{\tau}/m_{\mu}$ enhancement of the $S-\tau$ coupling, but is still not competitive with the other probes discussed above. Nevertheless, we show the corresponding sensitivity in Figs.~\ref{fig:money_plot} and~\ref{fig:money_plot_2}.

\subsection{Other fixed target and beam dump experiments}

Fixed target experiments with electron beams can be sensitive to the coupling $g_e$ through the process $e^- N \rightarrow e^- N S(\rightarrow \overline\chi \chi)$ in the lepton-specific model. This is analogous to the production mechanism discussed in~\secref{NA64}, except with the scalar being radiated off of an electron beam. Currently, the most sensitive of these experiments is NA64 running in electron mode. We re-interpret NA64-$e$ limits on the dark photon kinetic mixing parameter $\varepsilon$~\cite{Battaglieri:2017aum} as limits on an effective $S$-electron coupling $\varepsilon_S \equiv g_e/e$ where $e=\sqrt{4\pi \alpha}$.  The resulting bound is weaker than e.g.~bounds from current dark matter direct detection, and is not able to probe the region favored by the $(g-2)_\mu$ measurement. Of course, this conclusion could change if $g_e/g_\mu > m_e/m_\mu$.

Figs.~\ref{fig:money_plot}-\ref{fig:money_plot_2} show the expected reach of the proposed LDMX experiment~\cite{Berlin:2018bsc} with a 16 GeV electron beam. Here we translate the bound on $y\equiv \epsilon_\varphi^2 \alpha_\chi (m_\chi/m_S)^3$ from~\cite{Berlin:2018bsc} with $\epsilon_\varphi \equiv g_e /e$ into a bound on $g_\mu$. The expected reach of LDMX is impressive, but again, because of the coupling hierarchy, accessing the $a_{\mu}$-favored region for $m_S\gtrsim 300$ MeV in these models will require a muon beam experiment such as NA64-$\mu$ or M$^3$. A related search strategy is implemented in the proposed BDX experiment~\cite{Battaglieri:2016ggd}, which will be sensitive to the production of $\overline{\chi} \chi$ through $S$ and the subsequent scattering of the dark matter with a detector volume down-beam from the interaction point. The sensitivity of BDX to the lepton-specific model is translated from the corresponding dark photon projections~\cite{Battaglieri:2017aum} and shown Figs.~\ref{fig:money_plot}-\ref{fig:money_plot_2}. We use $E_\chi \approx E_\text{beam} = 11 \gev$ and $E_{e^-} \approx E_\text{th} = 350 \mev$~\cite{Battaglieri:2016ggd} in accounting for the amplitude difference between the scattering of the ultra-relativistic DM and electron via a scalar and vector-mediator (see also~\cite{Berlin:2018bsc}).The projected BDX sensitivity falls within the $(g-2)_\mu$ excluded regions and does not  exceed that of e.g.~LDMX. Analogous constraints from past beam dump experiments, such as E137~\cite{Bjorken:1988as, Batell:2014mga}, yield significantly weaker sensitivity and are not shown in Figs.~\ref{fig:money_plot}-\ref{fig:money_plot_2}.

\subsection{Summary}

To summarize, in the muon-specific case and for couplings near the size required to explain the $(g-2)_\mu$ discrepancy, none of above constraints are stronger than those from the $a_\mu$ measurements. This is reflected in the left-hand panels of Figs.~\ref{fig:money_plot} and~\ref{fig:money_plot_2}. Supernova cooling places an additional constraint on low masses and small couplings. Muon beam experiments, such as NA64-$\mu$, will thus be critical in exploring the parameter space of these models motivated by the $a_\mu$ discrepancy and/or thermal relic dark matter, absent significant fine-tuning. 

The right-hand panels of Figs.~\ref{fig:money_plot} and~\ref{fig:money_plot_2} compare the impact of the most sensitive aforementioned probes to the reach afforded by NA64-$\mu$ in models where the mediator couples to electrons, muons, and taus. Of the former, the most sensitive are cooling bounds from SN 1987A, which already constrain low masses/small couplings, and future mono-photon searches at Belle-II or Tera-Z factories. Future direct detection experiments will be sensitive to masses below $\sim 100$ MeV, and could, together with NA64-$\mu$, provide conclusive evidence for light leptophilic dark matter. LDMX-type electron fixed target experiments can also be sensitive to low masses, but by far the most powerful probe of these models would be provided by muon missing momentum experiments at NA64-$\mu$ or M$^3$.

\section{Conclusion}
\label{sec:Conclusion}

We have argued that sub-GeV dark matter with light leptophilic scalar mediators is a compelling target for current and future experiments. In addition to providing a viable dark matter candidate, these models can explain the long-standing $(g-2)_\mu$ discrepancy. Although not a UV complete scenario, light leptophilic dark matter can arise as a viable effective field theory with phenomenology that deviates in important ways from e.g.~gauged $L_\mu-L_\tau$ models. In particular, neutrino experimental constraints are absent, allowing for a larger range of masses consistent with $(g-2)_\mu$ and not currently ruled out by other experiments.  

Light DM with leptophilic scalar mediators is generally difficult to test, but we have argued that missing momentum searches at muon fixed target experiments, such as NA64-$\mu$, can provide valuable coverage of these scenarios. In particular, NA64-$\mu$ will be sensitive to the entire region of parameter space consistent with the measured value of $a_{\mu}$ for mediator masses below $\sim 5$ GeV. Even better sensitivity could be achieved utilizing the``1 vs 2'' technique for background elimination to accommodate 10$^{13}$ muons on target in a background-free environment. In addition to the $a_\mu$-favored regions, NA64-$\mu$ would also be able to explore a significant portion of the parameter space consistent with the observed dark matter density without fine-tuning. As such, muon beam experiments may afford us a first glimpse at what lies beyond the Standard Model.

\vspace{0.5cm}
\textbf{Note added:} As this paper was being finalized, Ref.~\cite{Berlin:2018bsc} appeared that also discusses searches for invisibly-decaying leptophilic scalar mediators at missing momentum experiments, and thus overlaps with some of our results and discussion above.

\section*{Acknowledgements}
We thank Dipanwita Banerjee, Marco Battaglieri, Asher Berlin, Jae Hyeok Chang, Patrick Draper, Bertrand Echenard, Rouven Essig, Stefania Gori, Eder Izaguirre, Yonatan Kahn, Simon Knapen, Gordan Krnjaic, Tongyan Lin, Samuel McDermott, David McKeen, Maxim Pospelov, Harikrishnan Ramani, Adam Riz, Martin Schmaltz, Natalia Toro, and Michael Williams for very helpful discussions. Part of this work was performed at the Aspen Center for Physics, which is supported by National Science Foundation grant PHY-1607611.
Y.Z. also would like to thank the KITP at UCSB for its hospitality during the completion of this work. 
The work of C.-Y.C is supported by NSERC, Canada. 
Research at the Perimeter Institute is supported in part by the Government of Canada through NSERC and by the Province of Ontario through MEDT. Y.Z. is supported through DOE grant DE-SC0015845. \appendix

\section{Scalar couplings to photons}
\label{sec:app}
In this appendix, we discuss the effective coupling of $S$ to photons relevant for computing the diphoton decay width, the $s$-channel contribution to $e^+ e^- \to S \gamma$ (represented by Fig.~\ref{fig:collider2}), and other $g_{\gamma\gamma}$-induced processes. It can be implemented in MC generator as a form factor. The coupling $g_{\gamma \gamma}$ is induced by muon loops in the muon-specific case, as well as electron and tau loops in the lepton-specific case. The computation proceeds similarly to the calculation of the Higgs-diphoton effective interaction in the SM, except allowing for general values of $p_S^2$ and $q^2$, the four-momentum--squared of $S$ and one of the photons, respectively (the other photon is assumed to be on-shell). We compute the lepton loop diagram using \texttt{Package-X 2.1}~\cite{Patel:2015tea} and match onto the tree-level effective coupling presented in \eqref{Lphoton}. We obtain the following result for the form factor $F_{1/2}$ defined in \eqref{Lphoton}:
\beq
\begin{aligned}
F_{1/2}(\tau,\rho) = &\f{2\tau}{\left(\rho \tau -1\right)^2}\left\{\rho \tau -1 + \sqrt{\rho(\rho-1)} \tau \log\left(1-2\rho+2\sqrt{\rho(\rho-1)}\right) + \right.\\
& \left. \f{1}{4}\left(\tau +\rho \tau -1 \right)\log^2\left(1-2\rho+2\sqrt{\rho(\rho-1)}\right) - \right. \\
&\left. \rho \left|\tau\right| \sqrt{1-\tau} \log \left(1-\f{2}{\tau}+\f{ 2\sqrt{1-\tau}}{\left|\tau\right|} \right) - \f{\tau + \rho \tau -1}{4}\log^2 \left(1-\f{2}{\tau}+\f{ 2\sqrt{1-\tau}}{\left|\tau\right|} \right) \right\},
\end{aligned}
\eeq
where 
\beq
\tau \equiv \f{4m_\ell^2}{p_S^2} ,\qquad \rho \equiv \f{q^2}{4 m_\ell^2} .
\eeq
The expression above is calculated using the $(+,-,-,-)$ metric convention. For computing the contribution to $e^+ e^- \to S \gamma$, $p_S^2 = m_S^2$ and $q^2 = s$, where $s$ is the center-of-mass energy of the given collider, $s \simeq (10.4 \, {\rm GeV})^2$, for BaBar, and $s\simeq (10.58 \, {\rm GeV})^2$ for Belle-II.

\begin{figure}[htbp]
   \centering
   \includegraphics[width=0.6\textwidth]{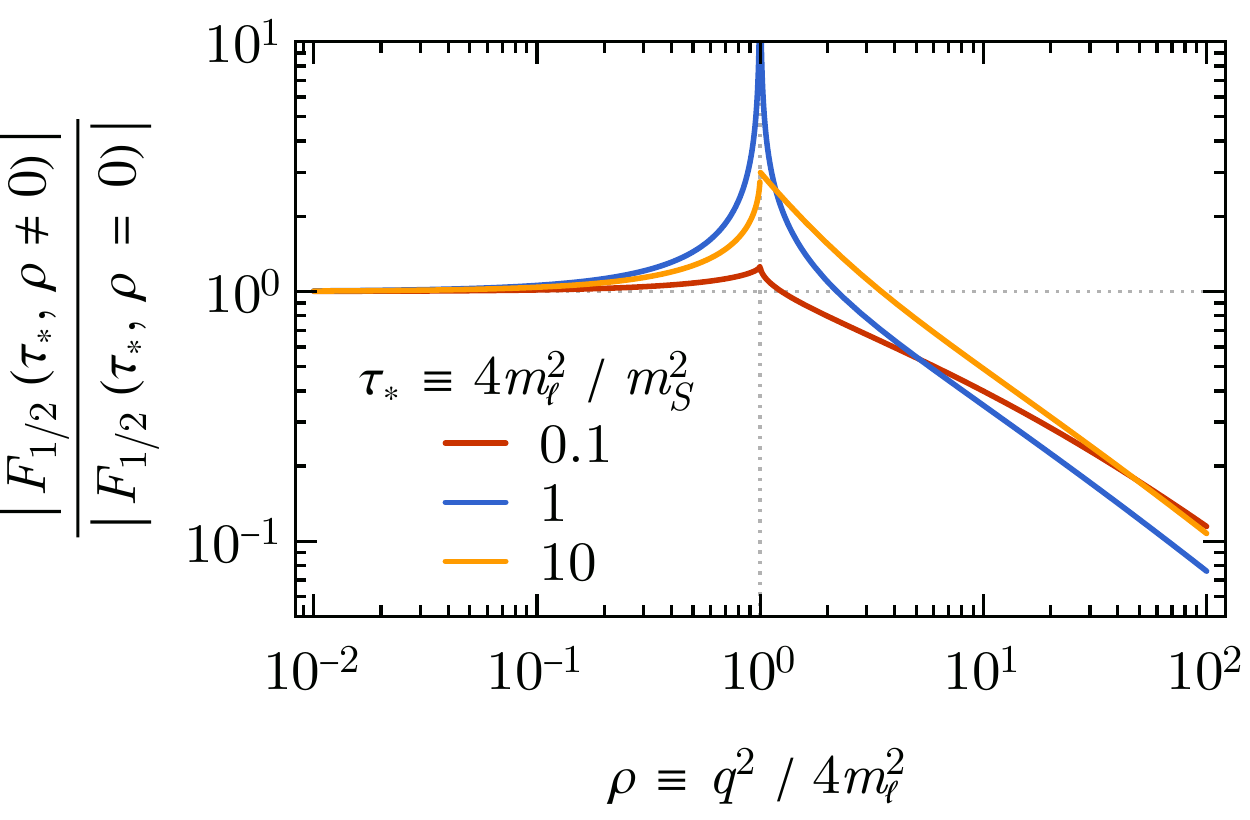} 
   \caption{The ratio of the absolute values of the $S\gamma \gamma$ loop function $F_{1/2}$ with an off-shell ($\rho\neq 0$) and on-shell ($\rho =0$) photon leg for a various values of $\tau$. The scalar and the other photon are taken to be on-shell.}
   \label{fig:loopfunction}
\end{figure}

In computing the $S\to \gamma \gamma$ decay width, $p_S^2 = m_S^2$ and $q^2=0$, which leads to the well-known result
\beq
F_{1/2} (\tau,0)=\left \{
  \begin{tabular}{l}
     $-2 \tau \left[1+\left(1-\tau\right) \left(\arcsin \tau^{-1/2}\right)^2\right]$ \quad $\tau \geq 1$\\\\
      $-2\tau \left[1-\f{1-\tau}{4}\left(-i \pi +\log\f{1+\sqrt{1-\tau}}{1-\sqrt{1-\tau}}\right)^2\right]$ \quad $\tau <1$.
  \end{tabular}
\right.
\eeq
\figref{loopfunction} shows the ratio of $|F_{1/2} (\tau, \rho\neq 0)|$ with respect to $|F_{1/2} (\tau, \rho= 0)|$ as a function of $\rho$ for fixed $\tau$. For $\rho \gg 1$, the loop function $F_{1/2} (\tau, \rho\neq 0)$ for various values of $\tau$ are significantly suppressed with respect to the on-shell loop function $F_{1/2}(\tau, 0)$.

\bibliography{Leptophilic_DM_at_NA64}

\providecommand{\href}[2]{#2}\begingroup\raggedright\begin{thebibliography}{10}

\bibitem{Olive:2016xmw}
{\scshape Particle Data Group} collaboration, C.~Patrignani et~al.,
  \emph{{Review of Particle Physics}},
  \href{http://dx.doi.org/10.1088/1674-1137/40/10/100001}{\emph{Chin. Phys.}
  {\bf C40} (2016) 100001}.

\bibitem{Bennett:2004pv}
{\scshape Muon g-2} collaboration, G.~W. Bennett et~al., \emph{{Measurement of
  the negative muon anomalous magnetic moment to 0.7 ppm}},
  \href{http://dx.doi.org/10.1103/PhysRevLett.92.161802}{\emph{Phys. Rev.
  Lett.} {\bf 92} (2004) 161802},
  [\href{http://arxiv.org/abs/hep-ex/0401008}{{\tt hep-ex/0401008}}].

\bibitem{Grange:2015fou}
{\scshape Muon g-2} collaboration, J.~Grange et~al., \emph{{Muon (g-2)
  Technical Design Report}},  \href{http://arxiv.org/abs/1501.06858}{{\tt
  1501.06858}}.

\bibitem{Anastasi:2015oea}
A.~Anastasi, \emph{{The Muon g-2 experiment at Fermilab}},
  \href{http://dx.doi.org/10.1051/epjconf/20159601002}{\emph{EPJ Web Conf.}
  {\bf 96} (2015) 01002}.

\bibitem{Mibe:2010zz}
{\scshape J-PARC g-2} collaboration, T.~Mibe, \emph{{New g-2 experiment at
  J-PARC}}, \href{http://dx.doi.org/10.1088/1674-1137/34/6/022}{\emph{Chin.
  Phys.} {\bf C34} (2010) 745--748}.

\bibitem{Nyffeler:2017ohp}
A.~Nyffeler, \emph{{Hadronic light-by-light scattering in the muon g-2}},  in
  \emph{{International Workshop on e+e- Collisions from Phi to Psi (PHIPSI17)
  Mainz, Germany, June 26-29, 2017}}, 2017.
\newblock \href{http://arxiv.org/abs/1710.09742}{{\tt 1710.09742}}.

\bibitem{Alexander:2016aln}
J.~Alexander et~al., \emph{{Dark Sectors 2016 Workshop: Community Report}},
  2016.
\newblock \href{http://arxiv.org/abs/1608.08632}{{\tt 1608.08632}}.

\bibitem{Battaglieri:2017aum}
M.~Battaglieri et~al., \emph{{US Cosmic Visions: New Ideas in Dark Matter 2017:
  Community Report}},  \href{http://arxiv.org/abs/1707.04591}{{\tt
  1707.04591}}.

\bibitem{Lee:1977ua}
B.~W. Lee and S.~Weinberg, \emph{{Cosmological Lower Bound on Heavy Neutrino
  Masses}}, \href{http://dx.doi.org/10.1103/PhysRevLett.39.165}{\emph{Phys.
  Rev. Lett.} {\bf 39} (1977) 165--168}.

\bibitem{Boehm:2002yz}
C.~Boehm, T.~A. Ensslin and J.~Silk, \emph{{Can Annihilating dark matter be
  lighter than a few GeVs?}},
  \href{http://dx.doi.org/10.1088/0954-3899/30/3/004}{\emph{J. Phys.} {\bf G30}
  (2004) 279--286}, [\href{http://arxiv.org/abs/astro-ph/0208458}{{\tt
  astro-ph/0208458}}].

\bibitem{Boehm:2003hm}
C.~Boehm and P.~Fayet, \emph{{Scalar dark matter candidates}},
  \href{http://dx.doi.org/10.1016/j.nuclphysb.2004.01.015}{\emph{Nucl. Phys.}
  {\bf B683} (2004) 219--263}, [\href{http://arxiv.org/abs/hep-ph/0305261}{{\tt
  hep-ph/0305261}}].

\bibitem{Pospelov:2007mp}
M.~Pospelov, A.~Ritz and M.~B. Voloshin, \emph{{Secluded WIMP Dark Matter}},
  \href{http://dx.doi.org/10.1016/j.physletb.2008.02.052}{\emph{Phys. Lett.}
  {\bf B662} (2008) 53--61}, [\href{http://arxiv.org/abs/0711.4866}{{\tt
  0711.4866}}].

\bibitem{ArkaniHamed:2008qn}
N.~Arkani-Hamed, D.~P. Finkbeiner, T.~R. Slatyer and N.~Weiner, \emph{{A Theory
  of Dark Matter}},
  \href{http://dx.doi.org/10.1103/PhysRevD.79.015014}{\emph{Phys. Rev.} {\bf
  D79} (2009) 015014}, [\href{http://arxiv.org/abs/0810.0713}{{\tt
  0810.0713}}].

\bibitem{Krnjaic:2015mbs}
G.~Krnjaic, \emph{{Probing Light Thermal Dark-Matter With a Higgs Portal
  Mediator}}, \href{http://dx.doi.org/10.1103/PhysRevD.94.073009}{\emph{Phys.
  Rev.} {\bf D94} (2016) 073009}, [\href{http://arxiv.org/abs/1512.04119}{{\tt
  1512.04119}}].

\bibitem{Gninenko:2014pea}
S.~N. Gninenko, N.~V. Krasnikov and V.~A. Matveev, \emph{{Muon g-2 and searches
  for a new leptophobic sub-GeV dark boson in a missing-energy experiment at
  CERN}}, \href{http://dx.doi.org/10.1103/PhysRevD.91.095015}{\emph{Phys. Rev.}
  {\bf D91} (2015) 095015}, [\href{http://arxiv.org/abs/1412.1400}{{\tt
  1412.1400}}].

\bibitem{Krasnikov:2017sma}
N.~V. Krasnikov, \emph{{Light scalars, ($g_{\mu}-2$) muon anomaly and dark
  matter in a model with a Higgs democracy}},
  \href{http://arxiv.org/abs/1707.00508}{{\tt 1707.00508}}.

\bibitem{Gninenko:2018tlp}
S.~N. Gninenko and N.~V. Krasnikov, \emph{{Probing the muon $g_\mu-2$ anomaly,
  $L_{\mu} - L_{\tau}$ gauge boson and Dark Matter in dark photon
  experiments}},  \href{http://arxiv.org/abs/1801.10448}{{\tt 1801.10448}}.

\bibitem{adeva1994measurement}
B.~Adeva, S.~Ahmad, A.~Arvidson, B.~Badelek, M.~Ballintijn, G.~Bardin et~al.,
  \emph{Measurement of the polarisation of a high energy muon beam},
  {\emph{Nuclear Instruments and Methods in Physics Research Section A:
  Accelerators, Spectrometers, Detectors and Associated Equipment} {\bf 343}
  (1994) 363--373}.

\bibitem{spsweb}
http://sba.web.cern.ch/sba/BeamsAndAreas/M2/M2-OperatorCourse.pdf.

\bibitem{Kahn:2018cqs}
Y.~Kahn, G.~Krnjaic, N.~Tran and A.~Whitbeck, \emph{{M$^3$: A New Muon Missing
  Momentum Experiment to Probe $(g-2)_{\mu}$ and Dark Matter at Fermilab}},
  \href{http://arxiv.org/abs/1804.03144}{{\tt 1804.03144}}.

\bibitem{Batell:2017kty}
B.~Batell, A.~Freitas, A.~Ismail and D.~Mckeen, \emph{{Flavor-specific scalar
  mediators}},  \href{http://arxiv.org/abs/1712.10022}{{\tt 1712.10022}}.

\bibitem{Batell:2016ove}
B.~Batell, N.~Lange, D.~McKeen, M.~Pospelov and A.~Ritz, \emph{{Muon anomalous
  magnetic moment through the leptonic Higgs portal}},
  \href{http://dx.doi.org/10.1103/PhysRevD.95.075003}{\emph{Phys. Rev.} {\bf
  D95} (2017) 075003}, [\href{http://arxiv.org/abs/1606.04943}{{\tt
  1606.04943}}].

\bibitem{Chen:2015vqy}
C.-Y. Chen, H.~Davoudiasl, W.~J. Marciano and C.~Zhang, \emph{{Implications of
  a Light "Dark Higgs" Solution to the $g_\mu-2$ Discrepancy}},
  \href{http://arxiv.org/abs/1511.04715}{{\tt 1511.04715}}.

\bibitem{Knapen:2017xzo}
S.~Knapen, T.~Lin and K.~M. Zurek, \emph{{Light Dark Matter: Models and
  Constraints}},
  \href{http://dx.doi.org/10.1103/PhysRevD.96.115021}{\emph{Phys. Rev.} {\bf
  D96} (2017) 115021}, [\href{http://arxiv.org/abs/1709.07882}{{\tt
  1709.07882}}].

\bibitem{Chen:2017awl}
C.-Y. Chen, M.~Pospelov and Y.-M. Zhong, \emph{{Muon Beam Experiments to Probe
  the Dark Sector}},
  \href{http://dx.doi.org/10.1103/PhysRevD.95.115005}{\emph{Phys. Rev.} {\bf
  D95} (2017) 115005}, [\href{http://arxiv.org/abs/1701.07437}{{\tt
  1701.07437}}].

\bibitem{Leveille:1977rc}
J.~P. Leveille, \emph{{The Second Order Weak Correction to (G-2) of the Muon in
  Arbitrary Gauge Models}},
  \href{http://dx.doi.org/10.1016/0550-3213(78)90051-2}{\emph{Nucl. Phys.} {\bf
  B137} (1978) 63--76}.

\bibitem{Lindner:2016bgg}
M.~Lindner, M.~Platscher and F.~S. Queiroz, \emph{{A Call for New Physics : The
  Muon Anomalous Magnetic Moment and Lepton Flavor Violation}},
  \href{http://dx.doi.org/10.1016/j.physrep.2017.12.001}{\emph{Phys. Rept.}
  {\bf 731} (2018) 1--82}, [\href{http://arxiv.org/abs/1610.06587}{{\tt
  1610.06587}}].

\bibitem{Jegerlehner:2018zrj}
F.~Jegerlehner, \emph{{The Muon g-2 in Progress}},
  \href{http://arxiv.org/abs/1804.07409}{{\tt 1804.07409}}.

\bibitem{Ambrogi:2018jqj}
F.~Ambrogi, C.~Arina, M.~Backovic, J.~Heisig, F.~Maltoni, L.~Mantani et~al.,
  \emph{{MadDM v.3.0: a Comprehensive Tool for Dark Matter Studies}},
  \href{http://arxiv.org/abs/1804.00044}{{\tt 1804.00044}}.

\bibitem{Alloul:2013bka}
\emph{{FeynRules 2.0 - A complete toolbox for tree-level phenomenology}},
  \href{http://dx.doi.org/10.1016/j.cpc.2014.04.012}{\emph{Comput. Phys.
  Commun.} {\bf 185} (2014) 2250--2300},
  [\href{http://arxiv.org/abs/1310.1921}{{\tt 1310.1921}}].

\bibitem{Feng:2017drg}
J.~L. Feng and J.~Smolinsky, \emph{{Impact of a resonance on thermal targets
  for invisible dark photon searches}},
  \href{http://dx.doi.org/10.1103/PhysRevD.96.095022}{\emph{Phys. Rev.} {\bf
  D96} (2017) 095022}, [\href{http://arxiv.org/abs/1707.03835}{{\tt
  1707.03835}}].

\bibitem{Ade:2015xua}
{\scshape Planck} collaboration, P.~A.~R. Ade et~al., \emph{{Planck 2015
  results. XIII. Cosmological parameters}},
  \href{http://dx.doi.org/10.1051/0004-6361/201525830}{\emph{Astron.
  Astrophys.} {\bf 594} (2016) A13},
  [\href{http://arxiv.org/abs/1502.01589}{{\tt 1502.01589}}].

\bibitem{Tulin:2013teo}
S.~Tulin, H.-B. Yu and K.~M. Zurek, \emph{{Beyond Collisionless Dark Matter:
  Particle Physics Dynamics for Dark Matter Halo Structure}},
  \href{http://dx.doi.org/10.1103/PhysRevD.87.115007}{\emph{Phys. Rev.} {\bf
  D87} (2013) 115007}, [\href{http://arxiv.org/abs/1302.3898}{{\tt
  1302.3898}}].

\bibitem{Peter:2012vi}
A.~H.~G. Peter, M.~Rocha, J.~S. Bullock and M.~Kaplinghat, \emph{{Cosmological
  Simulations with Self-Interacting Dark Matter II: Halo Shapes vs.
  Observations}}, {\emph{arXiv.org} (Aug., 2012) },
  [\href{http://arxiv.org/abs/1208.3026}{{\tt 1208.3026}}].

\bibitem{Balberg:2002ue}
S.~Balberg, S.~L. Shapiro and S.~Inagaki, \emph{{Selfinteracting Dark Matter
  Halos and the Gravothermal Catastrophe}},
  \href{http://dx.doi.org/10.1086/339038}{\emph{Astrophys. J.} {\bf 568} (2002)
  475--487}, [\href{http://arxiv.org/abs/astro-ph/0110561}{{\tt
  astro-ph/0110561}}].

\bibitem{Koda:2011yb}
J.~Koda and P.~R. Shapiro, \emph{{Gravothermal collapse of isolated
  self-interacting dark matter haloes: N-body simulation versus the fluid
  model}}, \href{http://dx.doi.org/10.1111/j.1365-2966.2011.18684.x}{\emph{Mon.
  Not. Roy. Astron. Soc.} {\bf 415} (2011) 1125},
  [\href{http://arxiv.org/abs/1101.3097}{{\tt 1101.3097}}].

\bibitem{Tulin:2017ara}
S.~Tulin and H.-B. Yu, \emph{{Dark Matter Self-interactions and Small Scale
  Structure}},  \href{http://arxiv.org/abs/1705.02358}{{\tt 1705.02358}}.

\bibitem{Cheng_1974}
T.~P. Cheng, E.~Eichten and L.-F. Li, \emph{Higgs phenomena in asymptotically
  free gauge theories},
  \href{http://dx.doi.org/10.1103/physrevd.9.2259}{\emph{Physical Review D}
  {\bf 9} (Apr, 1974) 2259--2273}.

\bibitem{cline:1998}
J.~M. Cline, ``Physics 198-730b: Quantum field theory.''
  http://www.physics.mcgill.ca/~jcline/qft1b.pdf.

\bibitem{deNiverville:2012ij}
P.~deNiverville, D.~McKeen and A.~Ritz, \emph{{Signatures of sub-GeV dark
  matter beams at neutrino experiments}},
  \href{http://dx.doi.org/10.1103/PhysRevD.86.035022}{\emph{Phys. Rev.} {\bf
  D86} (2012) 035022}, [\href{http://arxiv.org/abs/1205.3499}{{\tt
  1205.3499}}].

\bibitem{Slatyer:2015jla}
T.~R. Slatyer, \emph{{Indirect dark matter signatures in the cosmic dark ages.
  I. Generalizing the bound on s-wave dark matter annihilation from Planck
  results}}, \href{http://dx.doi.org/10.1103/PhysRevD.93.023527}{\emph{Phys.
  Rev.} {\bf D93} (2016) 023527}, [\href{http://arxiv.org/abs/1506.03811}{{\tt
  1506.03811}}].

\bibitem{Hardy:2016kme}
E.~Hardy and R.~Lasenby, \emph{{Stellar cooling bounds on new light particles:
  plasma mixing effects}},
  \href{http://dx.doi.org/10.1007/JHEP02(2017)033}{\emph{JHEP} {\bf 02} (2017)
  033}, [\href{http://arxiv.org/abs/1611.05852}{{\tt 1611.05852}}].

\bibitem{Gninenko:2001hx}
S.~N. Gninenko and N.~V. Krasnikov, \emph{{The Muon anomalous magnetic moment
  and a new light gauge boson}},
  \href{http://dx.doi.org/10.1016/S0370-2693(01)00693-1}{\emph{Phys. Lett.}
  {\bf B513} (2001) 119}, [\href{http://arxiv.org/abs/hep-ph/0102222}{{\tt
  hep-ph/0102222}}].

\bibitem{Kim:1973he}
K.~J. Kim and Y.-S. Tsai, \emph{{IMPROVED WEIZSACKER-WILLIAMS METHOD AND ITS
  APPLICATION TO LEPTON AND W BOSON PAIR PRODUCTION}},
  \href{http://dx.doi.org/10.1103/PhysRevD.8.3109}{\emph{Phys. Rev.} {\bf D8}
  (1973) 3109}.

\bibitem{Liu:2016mqv}
Y.-S. Liu, D.~McKeen and G.~A. Miller, \emph{{The Validity of the
  Weizsacker-Williams Approximation and the Analysis of Beam Dump
  Experiments}},  \href{http://arxiv.org/abs/1609.06781}{{\tt 1609.06781}}.

\bibitem{Gninenko:2017yus}
S.~N. Gninenko, D.~V. Kirpichnikov, M.~M. Kirsanov and N.~V. Krasnikov,
  \emph{{The exact tree-level calculation of the dark photon production in
  high-energy electron scattering at the CERN SPS}},
  \href{http://dx.doi.org/10.1016/j.physletb.2018.05.010}{\emph{Phys. Lett.}
  {\bf B782} (2018) 406--411}, [\href{http://arxiv.org/abs/1712.05706}{{\tt
  1712.05706}}].

\bibitem{pdgweb}
http://pdg.lbl.gov/2017/AtomicNuclearProperties/HTML/lead\_Pb.html.

\bibitem{na64web}
https://na64.web.cern.ch/content/muon-g-2-and-new-leptonic-dark-boson.

\bibitem{Dolan:2017osp}
M.~J. Dolan, T.~Ferber, C.~Hearty, F.~Kahlhoefer and K.~Schmidt-Hoberg,
  \emph{{Revised constraints and Belle II sensitivity for visible and invisible
  axion-like particles}},
  \href{http://dx.doi.org/10.1007/JHEP12(2017)094}{\emph{JHEP} {\bf 12} (2017)
  094}, [\href{http://arxiv.org/abs/1709.00009}{{\tt 1709.00009}}].

\bibitem{Berlin:2018bsc}
A.~Berlin, N.~Blinov, G.~Krnjaic, P.~Schuster and N.~Toro, \emph{{Dark Matter,
  Millicharges, Axion and Scalar Particles, Gauge Bosons, and Other New Physics
  with LDMX}},  \href{http://arxiv.org/abs/1807.01730}{{\tt 1807.01730}}.

\bibitem{Liu:2017zdh}
J.~Liu, L.-T. Wang, X.-P. Wang and W.~Xue, \emph{{Exposing the dark sector with
  future Z factories}},
  \href{http://dx.doi.org/10.1103/PhysRevD.97.095044}{\emph{Phys. Rev.} {\bf
  D97} (2018) 095044}, [\href{http://arxiv.org/abs/1712.07237}{{\tt
  1712.07237}}].

\bibitem{Bollig:2017lki}
R.~Bollig, H.~T. Janka, A.~Lohs, G.~Martinez-Pinedo, C.~J. Horowitz and
  T.~Melson, \emph{{Muon Creation in Supernova Matter Facilitates
  Neutrino-driven Explosions}},
  \href{http://dx.doi.org/10.1103/PhysRevLett.119.242702}{\emph{Phys. Rev.
  Lett.} {\bf 119} (2017) 242702}, [\href{http://arxiv.org/abs/1706.04630}{{\tt
  1706.04630}}].

\bibitem{Raffelt:1996wa}
G.~G. Raffelt, \emph{{Stars as laboratories for fundamental physics}}.
\newblock Chicago University Press, 1996.

\bibitem{Chang:2018rso}
J.~H. Chang, R.~Essig and S.~D. McDermott, \emph{{Supernova 1987A Constraints
  on Sub-GeV Dark Sectors, Millicharged Particles, the QCD Axion, and an
  Axion-like Particle}},  \href{http://arxiv.org/abs/1803.00993}{{\tt
  1803.00993}}.

\bibitem{inprogress}
J.~H. Chang and H.~Ramani, \emph{{Supernova Constraints on Muonic Forces}}, to appear.

\bibitem{Essig:2011nj}
R.~Essig, J.~Mardon and T.~Volansky, \emph{{Direct Detection of Sub-GeV Dark
  Matter}}, \href{http://dx.doi.org/10.1103/PhysRevD.85.076007}{\emph{Phys.
  Rev.} {\bf D85} (2012) 076007}, [\href{http://arxiv.org/abs/1108.5383}{{\tt
  1108.5383}}].

\bibitem{Essig:2015cda}
R.~Essig, M.~Fernandez-Serra, J.~Mardon, A.~Soto, T.~Volansky and T.-T. Yu,
  \emph{{Direct Detection of sub-GeV Dark Matter with Semiconductor Targets}},
  \href{http://dx.doi.org/10.1007/JHEP05(2016)046}{\emph{JHEP} {\bf 05} (2016)
  046}, [\href{http://arxiv.org/abs/1509.01598}{{\tt 1509.01598}}].

\bibitem{Essig:2012yx}
R.~Essig, A.~Manalaysay, J.~Mardon, P.~Sorensen and T.~Volansky, \emph{{First
  Direct Detection Limits on sub-GeV Dark Matter from XENON10}},
  \href{http://dx.doi.org/10.1103/PhysRevLett.109.021301}{\emph{Phys. Rev.
  Lett.} {\bf 109} (2012) 021301}, [\href{http://arxiv.org/abs/1206.2644}{{\tt
  1206.2644}}].

\bibitem{Essig:2017kqs}
R.~Essig, T.~Volansky and T.-T. Yu, \emph{{New Constraints and Prospects for
  sub-GeV Dark Matter Scattering off Electrons in Xenon}},
  \href{http://dx.doi.org/10.1103/PhysRevD.96.043017}{\emph{Phys. Rev.} {\bf
  D96} (2017) 043017}, [\href{http://arxiv.org/abs/1703.00910}{{\tt
  1703.00910}}].

\bibitem{Angle:2011th}
{\scshape XENON10} collaboration, J.~Angle et~al., \emph{{A search for light
  dark matter in XENON10 data}},
  \href{http://dx.doi.org/10.1103/PhysRevLett.110.249901,
  10.1103/PhysRevLett.107.051301}{\emph{Phys. Rev. Lett.} {\bf 107} (2011)
  051301}, [\href{http://arxiv.org/abs/1104.3088}{{\tt 1104.3088}}].

\bibitem{Aprile:2016wwo}
{\scshape XENON} collaboration, E.~Aprile et~al., \emph{{Low-mass dark matter
  search using ionization signals in XENON100}},
  \href{http://dx.doi.org/10.1103/PhysRevD.94.092001,
  10.1103/PhysRevD.95.059901}{\emph{Phys. Rev.} {\bf D94} (2016) 092001},
  [\href{http://arxiv.org/abs/1605.06262}{{\tt 1605.06262}}].

\bibitem{Agnes:2018oej}
{\scshape DarkSide} collaboration, P.~Agnes et~al., \emph{{Constraints on
  Sub-GeV Dark Matter-Electron Scattering from the DarkSide-50 Experiment}},
  \href{http://arxiv.org/abs/1802.06998}{{\tt 1802.06998}}.

\bibitem{Agnese:2018col}
{\scshape SuperCDMS} collaboration, R.~Agnese et~al., \emph{{First Dark Matter
  Constraints from a SuperCDMS Single-Charge Sensitive Detector}},
  {\emph{Submitted to: Phys. Rev. Lett.} (2018) },
  [\href{http://arxiv.org/abs/1804.10697}{{\tt 1804.10697}}].

\bibitem{Chauhan:2016joa}
B.~Chauhan and S.~Mohanty, \emph{{Constraints on leptophilic light dark matter
  from internal heat flux of Earth}},
  \href{http://dx.doi.org/10.1103/PhysRevD.94.035024}{\emph{Phys. Rev.} {\bf
  D94} (2016) 035024}, [\href{http://arxiv.org/abs/1603.06350}{{\tt
  1603.06350}}].

\bibitem{TheBABAR:2016rlg}
{\scshape BaBar} collaboration, J.~P. Lees et~al., \emph{{Search for a muonic
  dark force at BABAR}},
  \href{http://dx.doi.org/10.1103/PhysRevD.94.011102}{\emph{Phys. Rev.} {\bf
  D94} (2016) 011102}, [\href{http://arxiv.org/abs/1606.03501}{{\tt
  1606.03501}}].

\bibitem{Babusci:2015zda}
{\scshape KLOE-2} collaboration, A.~Anastasi et~al., \emph{{Search for dark
  Higgsstrahlung in $ee \to \mu\mu$ and missing energy events with the KLOE
  experiment}},
  \href{http://dx.doi.org/10.1016/j.physletb.2015.06.015}{\emph{Phys. Lett.}
  {\bf B747} (2015) 365--372}, [\href{http://arxiv.org/abs/1501.06795}{{\tt
  1501.06795}}].

\bibitem{Artamonov:2016wby}
{\scshape E949} collaboration, A.~V. Artamonov et~al., \emph{{Search for the
  rare decay $K^+\to\mu^+\nu\bar\nu\nu$}},
  \href{http://dx.doi.org/10.1103/PhysRevD.94.032012}{\emph{Phys. Rev.} {\bf
  D94} (2016) 032012}, [\href{http://arxiv.org/abs/1606.09054}{{\tt
  1606.09054}}].

\bibitem{Berlin:2018pwi}
A.~Berlin, S.~Gori, P.~Schuster and N.~Toro, \emph{{Dark Sectors at the
  Fermilab SeaQuest Experiment}},  \href{http://arxiv.org/abs/1804.00661}{{\tt
  1804.00661}}.

\bibitem{Acciarri:1997im}
{\scshape L3} collaboration, M.~Acciarri et~al., \emph{{Search for new physics
  in energetic single photon production in $e^{+} e^{-}$ annihilation at the
  $Z$ resonance}},
  \href{http://dx.doi.org/10.1016/S0370-2693(97)01003-4}{\emph{Phys. Lett.}
  {\bf B412} (1997) 201--209}.

\bibitem{Gunion:1989we}
J.~F. Gunion, H.~E. Haber, G.~L. Kane and S.~Dawson, \emph{{The Higgs Hunter's
  Guide}}, {\emph{Front. Phys.} {\bf 80} (2000) 1--404}.

\bibitem{Djouadi:2005gi}
A.~Djouadi, \emph{{The Anatomy of electro-weak symmetry breaking. I: The Higgs
  boson in the standard model}},
  \href{http://dx.doi.org/10.1016/j.physrep.2007.10.004}{\emph{Phys. Rept.}
  {\bf 457} (2008) 1--216}, [\href{http://arxiv.org/abs/hep-ph/0503172}{{\tt
  hep-ph/0503172}}].

\bibitem{dEnterria:2016sca}
D.~d'Enterria, \emph{{Physics at the FCC-ee}},  in \emph{{Proceedings, 17th
  Lomonosov Conference on Elementary Particle Physics: Moscow, Russia, August
  20-26, 2015}}, pp.~182--191, 2017.
\newblock \href{http://arxiv.org/abs/1602.05043}{{\tt 1602.05043}}.
\newblock \href{http://dx.doi.org/10.1142/9789813224568_0028}{DOI}.

\bibitem{CEPC-SPPCStudyGroup:2015csa}
C.-S.~S. Group, \emph{{CEPC-SPPC Preliminary Conceptual Design Report. 1.
  Physics and Detector}}, .

\bibitem{Alwall:2014hca}
J.~Alwall, R.~Frederix, S.~Frixione, V.~Hirschi, F.~Maltoni, O.~Mattelaer
  et~al., \emph{{The automated computation of tree-level and next-to-leading
  order differential cross sections, and their matching to parton shower
  simulations}}, \href{http://dx.doi.org/10.1007/JHEP07(2014)079}{\emph{JHEP}
  {\bf 07} (2014) 079}, [\href{http://arxiv.org/abs/1405.0301}{{\tt
  1405.0301}}].

\bibitem{Battaglieri:2016ggd}
{\scshape BDX} collaboration, M.~Battaglieri et~al., \emph{{Dark Matter Search
  in a Beam-Dump eXperiment (BDX) at Jefferson Lab}},
  \href{http://arxiv.org/abs/1607.01390}{{\tt 1607.01390}}.

\bibitem{Bjorken:1988as}
J.~D. Bjorken, S.~Ecklund, W.~R. Nelson, A.~Abashian, C.~Church, B.~Lu et~al.,
  \emph{{Search for Neutral Metastable Penetrating Particles Produced in the
  SLAC Beam Dump}},
  \href{http://dx.doi.org/10.1103/PhysRevD.38.3375}{\emph{Phys. Rev.} {\bf D38}
  (1988) 3375}.

\bibitem{Batell:2014mga}
B.~Batell, R.~Essig and Z.~Surujon, \emph{{Strong Constraints on Sub-GeV Dark
  Sectors from SLAC Beam Dump E137}},
  \href{http://dx.doi.org/10.1103/PhysRevLett.113.171802}{\emph{Phys. Rev.
  Lett.} {\bf 113} (2014) 171802}, [\href{http://arxiv.org/abs/1406.2698}{{\tt
  1406.2698}}].

\bibitem{Patel:2015tea}
H.~H. Patel, \emph{{Package-X: A Mathematica package for the analytic
  calculation of one-loop integrals}},
  \href{http://dx.doi.org/10.1016/j.cpc.2015.08.017}{\emph{Comput. Phys.
  Commun.} {\bf 197} (2015) 276--290},
  [\href{http://arxiv.org/abs/1503.01469}{{\tt 1503.01469}}].

\end{thebibliography}\endgroup

\bibliographystyle{JHEP}

\end{document}